\documentclass[a4paper,11pt]{article}

\usepackage{jheppub} 

\usepackage[T1]{fontenc} 

\usepackage{xcolor}
\usepackage[utf8]{inputenc}
\usepackage{amsmath}
\usepackage{amssymb}
\usepackage{pifont}
\usepackage{graphicx}
\usepackage{multirow}
\usepackage[normalem]{ulem}
\usepackage{comment}
\usepackage{booktabs}
\usepackage{slashed}
\usepackage{makecell}

\usepackage{tikz-feynman}
\tikzfeynmanset{compat=1.1.0}
\usetikzlibrary{arrows,automata,positioning}
\tikzset{
  vertex/.style={circle,draw, minimum size=1.5em},
    edge/.style={->,> = latex'}
}
\tikzfeynmanset{
  fermion1/.style={
    /tikz/postaction={
      /tikz/decoration={
        markings,
        mark=at position 0.5 with {
          \node[
            transform shape,
            xshift=-0.5mm,
            fill,
            dart tail angle=100,
            inner sep=1.3pt,
            draw=none,
            dart
          ] { };
        },
      },
      /tikz/decorate=true,
    },
  },
  fermion2/.style={
    /tikz/postaction={
      /tikz/decoration={
        markings,
        mark=at position 0.5 with {
          \arrow{>[length=5pt, width=4pt]};
        },
      },
      /tikz/decorate=true,
    },
  },
}

\newcommand{\be}{\begin{eqnarray*}}
\newcommand{\ee}{\end{eqnarray*}}

\newcommand{\bee}{\begin{eqnarray}}
\newcommand{\eee}{\end{eqnarray}}
\newcommand{\beeq}{\begin{equation}}
\newcommand{\eeq}{\end{equation}}

\newcommand{\ba}{\begin{array}}
\newcommand{\ea}{\end{array}}
\newcommand{\bd}{\begin{displaymath}}
\newcommand{\ed}{\end{displaymath}}
\newcommand{\besub}{\begin{subequations}}
\newcommand{\eesub}{\end{subequations}}
\newcommand{\bea}{\begin{eqnarray}}
\newcommand{\eea}{\end{eqnarray}}

\def\bra{\langle}
\def\ket{\rangle}

\def\l{\lambda}

\def\L{\Lambda}

\def\q2 {q^2}

\def\r {\rightarrow}

\def\bt{\begin{table}}
\def\et{\end{table}}
\def\zf{\mathcal{Z}_4}

\newdimen\arrowsize
\newdimen\mylw
\pgfkeys{/my arrows/chemeq/.style={draw,thick,double distance=2pt,onearc-onearc}}
\pgfkeys{/my arrows/size/.code={\pgfsetarrowoptions{onearc}{#1}}}
\def\myalw{.4pt}
\pgfarrowsdeclare{onearc}{onearc}{%
  \mylw=\myalw
  \pgfarrowsleftextend{-\pgfgetarrowoptions{onearc}-.5\mylw}
  \pgfarrowsrightextend{0pt}
}{%
  \pgfsetdash{}{0pt}
  \mylw=\pgflinewidth
  \pgfsetlinewidth{\myalw}
  \advance\arrowsize by.5\pgflinewidth
  \pgfpathmoveto{\pgfpoint{-\pgfgetarrowoptions{onearc}}{-\pgfgetarrowoptions{onearc}-.5\mylw}}%
  \pgfpatharc{180}{90}{\pgfgetarrowoptions{onearc}}
  \pgfusepathqstroke
}

\title{\boldmath  
	 Leptogenesis, Dark Matter and Gravitational Waves from Discrete Symmetry Breaking}
  
\author[a]{Subhaditya Bhattacharya,}
\author[a]{Niloy Mondal,}
\author[b]{Rishav Roshan,}
\author[c,d]{and Drona Vatsyayan}

\affiliation[a]{Department of Physics, Indian Institute of Technology Guwahati, Assam 781039, India}

\affiliation[b]{School of Physics and Astronomy, University of Southampton, Southampton SO17 1BJ, United Kingdom}

\affiliation[c]{Departament de Física Teòrica, Universitat de València, 46100 Burjassot, Spain}

\affiliation[d]{Instituto de Física Corpuscular (CSIC-Universitat de València),
Parc Científic UV, C/Catedrático José Beltrán, 2, E-46980 Paterna, Spain}

\emailAdd{subhab@iitg.ac.in}
\emailAdd{niloy18@iitg.ac.in}
\emailAdd{r.roshan@soton.ac.uk}
\emailAdd{drona.vatsyayan@ific.uv.es}

\abstract{We analyse a model that connects the neutrino sector and the dark sector of the universe  via a mediator $\Phi$, stabilised by a discrete  $\mathcal{Z}_4$ symmetry that breaks to a remnant  $\mathcal{Z}_2$ upon $\Phi$ acquiring a non-zero
vacuum expectation value ($v_\phi$). The model accounts for the observed baryon asymmetry of the universe via additional contributions to the canonical Type-I leptogenesis. The $\mathcal{Z}_4$ symmetry breaking scale ($v_\phi$) in the model not only establishes a connection between the neutrino sector and the dark sector, but could also lead to gravitational wave signals that are within the reach of current and future experimental sensitivities.}


\makeatletter
\gdef\@fpheader{}
\makeatother

\begin{document} 
\maketitle
\flushbottom
\section{Introduction}

Numerous cosmological and experimental observations strongly 
suggest the presence of new physics (NP) beyond the 
Standard Model (BSM). Notably, the origin of tiny neutrino masses,
the cosmic baryon asymmetry along with the dark 
matter (DM) that makes up roughly 85\% of the matter content 
in our Universe, persist as open problems in particle physics 
despite numerous theoretical and experimental efforts to 
explain them. Absence of a concrete hint of BSM physics in ongoing 
particle physics experiments indicate the possibility of NP 
to exist at a high scale or weakly coupled, often parameterised via 
effective operators \cite{Buchmuller:1985jz,Grzadkowski:2010es,Masso:2014xra,Englert:2014cva,Calibbi:2017uvl,Aebischer:2021uvt,Azatov:2016sqh,Aguilar-Saavedra:2018ksv,CMS:2020lrr,Degrande:2012wf,Coito:2022kif,Beneito:2023xbk}, posing significant challenges to 
its testability. Consequently, alternative search strategies 
that extend beyond collider searches to probe these NP 
scenarios, like gravitational waves (GWs) by the 
LIGO-VIRGO collaboration \cite{LIGOScientific:2016aoc} 
have attracted considerable attention in recent times.

Amongst recent theoretical efforts, of particular interest are 
those suggesting minimal extensions to the Standard Model (SM) 
to address the seemingly unrelated sectors of neutrinos, DM 
and excess of baryonic matter over anti-baryonic one in a 
unified framework \cite{Bhattacharya:2018ljs, Asaka:2005an,Datta:2021elq,Bhattacharya:2021jli,Herrero-Garcia:2023lhv,Falkowski:2011xh,DuttaBanik:2020vfr,An:2009vq,Cosme:2005sb,Chun:2011cc,Barman:2021tgt,Ma:2006fn,Hambye:2006zn,Gu:2007ug,Gu:2008yj,Aoki:2008av,Aoki:2009vf,Josse-Michaux:2011sjn}. 
However, the requirement of tiny neutrino mass, a very specific 
DM relic density together with the fine-tuned matter-anti 
matter asymmetry leads naturally to heavy BSM physics or 
those weakly coupled to visible sector. Moreover, a framework 
where there is a connection between all these sectors 
(not just having them together under one umbrella), is not 
only theoretically motivating, but also the outcome of one sector 
is interestingly poised with the other sector, making it both
phenomenologically appealing and challenging at the same time. 

In this work, we will consider a seesaw I framework with a 
dimension five effective operator in presence of a real 
scalar singlet ($\Phi$), which also provides additional 
features to the vanilla leptogenesis. The same $\Phi$ 
acts as a mediator between the neutrino sector and a 
fermion DM, thus connecting the phenomenology of all three 
sectors. The connection is very much dependent on 
a discrete $\zf$ symmetry (there are other possibilities, but this 
choice is the simplest), which is broken by $\bra \Phi \ket$ 
into a $\mathcal{Z}_2$ symmetry to stabilize the DM candidate.
The model is motivated from \cite{Bhattacharya:2018ljs}, but 
has important distinction in phenomenological outcome. 

As the scrutiny around the weakly interacting massive particle 
(WIMP) freeze-out scenario continues and 
its parameter space becomes increasingly constrained, alternative
mechanisms for reproducing the observed dark matter relic abundance
$\Omega_{\rm DM}h^2 = 0.12 \pm 0.0012$ are being explored
\cite{ParticleDataGroup:2022pth}. Our interest lies
in the freeze-in mechanism \cite{Hall:2009bx}, where the 
DM candidate, a feebly interacting massive particle (FIMP), is 
mainly produced out-of-equilibrium, via its interactions with
particles in the thermal bath having feeble coupling 
$y \sim \mathcal{O}(10^{-10})$ and saves the DM
particle from direct/collider searches with possible exceptions. 

The adoption of a FIMP-like DM further necessitates a 
significantly large value of $v_\phi$ to generate a sizeable 
DM mass. This in turn pushes the scale of seesaw towards higher values, and accommodates baryogenesis via 
thermal leptogenesis \cite{Fukugita:1986hr}. Moreover, the model set-up also leads to novel scattering processes that can generate a CP asymmetry as well as lead to a washout of the decay-generated asymmetry. Scattering processes contributing to baryogenesis have been studied earlier in the literature, though, mostly involving particles from the hidden sector or dark matter \cite{Bento:2001rc, Gu:2009yx,Baldes:2014gca, Baldes:2014rda,Ghosh:2021ivn,Cui:2011ab, Bernal:2012gv, Bernal:2013bga, Kumar:2013uca, Baldes:2015lka,Racker:2014uga,Goudelis:2022bls,Dasgupta:2019lha}.

Additionally, the high-scale breaking of a discrete symmetry can result in the formation of two-dimensional sheet-like topological defects known as domain walls (DW) \cite{Saikawa:2017hiv,Roshan:2024qnv}. In general, such defects act as a cosmological catastrophe, as they can soon dominate the energy density of the Universe. This is because their energy density $\rho_{\text{DW}}$ scales as $a^{-1}$ ($a$ denotes the scale factor), thus resulting in a slower dilution rate compared to the energy density of matter or radiation. To avoid this cosmological catastrophe, one approach involves rendering the DW unstable, allowing for its collapse before it overcloses the Universe. This instability can be ensured if the discrete symmetry is only approximate and explicitly broken by a small parameter in the theory. In such a situation, the collision and annihilation of the DWs can produce a stochastic gravitational wave (GW) background that can be detected by the present and future GW detectors. This provides a window to not only the high-scale seesaw models but also to probe the feebly interacting dark sector, which we study here.

The paper is structured as follows, in Section~\ref{sec:model} we discuss the model, the scalar potential is studied in Section~\ref{sec:scalar}. The contribution to DM relic abundance and neutrino masses is discussed in Sections~\ref{sec:DM} and \ref{sec:numass}, respectively. We study the new scattering contributions to the leptogenesis mechanism and provide the numerical results in Section~\ref{sec:lepto}. In Section~\ref{sec:GW}, we discuss the production of gravitational waves and the prospects for testing the discrete symmetry breaking scale. We draw our conclusions in Section~\ref{sec:summary}.  

\section{The Model}\label{sec:model}

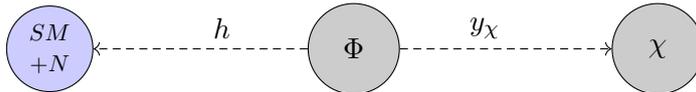
\begin{figure}[!htb]
    \centering
     \begin{tikzpicture}[scale=1.0, transform shape,every text node part/.style={align=center}]
        \node[draw,circle,fill=gray!40!white,minimum size=1.2cm,scale=1] (P0) at (0,0) {$\Phi$};
        \node[draw,circle,fill=blue!20!white,minimum size=1.2cm,scale=0.8] (P1) at (-4,0) {$SM$\\$+N$};
        \node[draw,circle,fill=gray!40!white,minimum size=1.2cm,scale=1] (P2) at (4,0) {$\chi$};
        \draw[densely dashed,->] (P0) -- (P1) node[pos=.4, above] {$h$};
        \draw[densely dashed,->] (P0) -- (P2) node[pos=.4, above] {$y_{\chi}$};   
    \end{tikzpicture}
    \caption{Schematic framework of the DM interaction with SM via the mediator $\Phi$. The gray (blue) blob denotes the dark (visible) sector.}
\label{fig:scheme}
\end{figure}

We introduce a real singlet scalar field $(\Phi)$, 
three right-handed neutrinos $(N)$ and a singlet Majorana 
fermion DM $(\chi)$ to the existing SM field content (similar to 
~\cite{Bhattacharya:2018ljs}). $\Phi$ serves as a mediator to 
connect the neutrino and dark sectors (see Fig.~\ref{fig:scheme}). 

The new fields are charged under a discrete $\mathcal{Z}_4$ symmetry (while the SM fields are uncharged). $\Phi$ with $\mathcal{Z}_4$ charge $q$ transforms as $\Phi \r e^{i\pi\,q/2}\Phi$, so do $N$ and $\chi$. Additionally, $N$ carries a lepton number $\text{LN}=-1$ where $\chi$ does not. The $\zf$ charges along with the quantum numbers of the new fields are shown in Table~\ref{tab:fields}. The Lagrangian respecting SM$\times \mathcal{Z}_4$ symmetry for the model can thus be written as
\bea \label{eq:lag}
    -\mathcal{L}\supset \frac{h_{\alpha i}}{\Lambda} \bar{l}_L^\alpha  \tilde{H}\Phi N_R^i + y_{\chi} \Phi\overline{\chi^c} \chi  + {M_N}_{ij}\overline{{N^c_R}^i}N_R^j + V(H,\Phi) \text{h.c.}\,,
\eea
where $\Lambda$ denotes the cut-off scale of the theory, $\{\alpha=e,\mu,\tau\}$ and $\{i,j=1,2,3\}$ denote the family  and generation indices, $l_L$ is the left-handed SM lepton doublet and $H$ is the SM Higgs iso-doublet $(\tilde{H}=i\sigma_2 H^\ast)$, and $V(H,\Phi)$ is the scalar potential that we discuss below. The coupling $h$ is a general $3 \times 3$ complex matrix, $M_N$ is a $3 \times 3$ complex symmetric matrix and $y_{\chi}$ is an extremely small real number (leading to the feeble interaction). The first term in Eq.~\eqref{eq:lag} is a non-renormalizable term that is allowed by both LN and dark sector symmetry and prohibits terms like $\bar{l}_L \tilde{H} N$, $\bar{N^c}N \Phi$ as well as Majorana mass for DM. The 
mass term for $N$ breaks LN symmetry, but is allowed by $\zf$ symmetry.
\begin{table}[!htb]
    \centering
    \begin{tabular}{c c c c c c c}
        \hline \hline
          & Field & Generations & \small{$SU(3)_C \times SU(2)_L \times U(1)_Y$} & $q$ & LN\\
          \hline
          \textbf{Scalar:} & $\Phi$ & 1 & $(1,1,0)$ & 2 & 0\\
          \multirow{2}{*}{\textbf{Fermions:}}& $\chi$ & 1 & $(1,1,0)$ & 1 & 0\\
          & $N$ & 3 & $(1,1,0)$ & 2 & -1\\
          \hline \hline
    \end{tabular}
    \caption{List of particles added and their charge assignments.}
    \label{tab:fields}
\end{table}

The connection between the two sectors becomes apparent once $\Phi$ acquires a non-zero vacuum expectation value (VEV), $\bra \Phi \ket = v_\phi$. On one hand, the term $\Phi\overline{\chi^c} \chi$ leads to massive DM with $M_\chi \propto v_\phi$,
\begin{equation}\label{eq:dmmass}
    M_\chi = 100{\rm~GeV}\, \left(\frac{y_\chi}{10^{-10}}\right)\, \left(\frac{v_\phi}{10^{12}}\right)\,,
\end{equation}
on the other, it leads to an effective Yukawa coupling for the neutrino sector ${y_\nu}_{\alpha i} \equiv (h_{\alpha i}v_\phi)/\Lambda$ to generate the light neutrino masses via the usual type-I seesaw mechanism
\begin{equation}\label{eq:numass}
{m_\nu} = h h^T\frac{v_\phi^2}{\Lambda^2} \frac{v^2}{M_N}=y_\nu y_\nu^T\, \frac{v^2 }{ M_N}\,,
\end{equation}
where $v=246$ GeV is the SM Higgs VEV. Note here the dependence on 
$\{v_\phi,\Lambda\}$ gives an overall numerical factor, which
is absorbed into $y_\nu$ and should be consistent with the 
neutrino oscillation data. 

It can be seen from Eq.~\eqref{eq:dmmass}, that in order to generate a sizeable mass for $\chi$, one requires a large $v_\phi$, 
given the feeble coupling $y_\chi$. This opens up a 
new phenomenological paradigm, given the dependence of both 
neutrino and the dark sector on this VEV, as opposed to \cite{Bhattacharya:2018ljs}, where $v_\phi \sim \mathcal{O}(10^2) {\rm~GeV}$ was considered. We should also note that one could 
write a dimension-5 term $\Phi^2 \bar{N^c}N/\Lambda$, 
consistent with the symmetries of the model, which can contribute 
to the light neutrino effective mass once $\Phi$ gets a 
VEV, however, such contribution can simply be taken into 
account by rewriting $M_N$, without any loss of generality.

\section{The scalar sector}\label{sec:scalar}

The full scalar potential involving the singlet $\Phi$ and SM Higgs consistent with the symmetries of the model can be written as
\bea \label{eq:scalarpot}
V(H,\Phi) &=& -\mu_H^2 H^{\dagger} H - \frac{\mu_{\phi}^2}{2} \Phi^2
+ \lambda_H (H^{\dagger} H)^2 + \frac{\lambda_{\phi}}{4}\Phi^4+\frac{\lambda_{H\phi}}{2}(H^{\dagger} H)\Phi^2\,,
\eea
where $\mu_\phi^2 > 0$ is chosen to trigger spontaneous symmetry breaking for $\Phi$. Note that once $\Phi$ acquires a large VEV, 
the discrete $\zf$ symmetry is broken to a remnant 
$\mathcal{Z}_2$ symmetry that stabilizes the dark matter 
candidate. After electroweak symmetry breaking (EWSB), 
the $CP$-even component of $H$ also receives VEV $v$. Then 
$v_{\phi}$ and $v$ are related through the tadpole conditions as,
\bea
\mu_H^2 = \l_H v^2 + \frac{\l_{H\phi}}{2} v_{\phi}^2\,,\quad
\mu_\phi^2 = \l_\phi v_{\phi}^2 + \frac{\l_{H\phi}}{2} v^2\,.
\eea 
The scalar multiplets can then be parameterised as,
\begin{equation}
    H = \frac{1}{\sqrt{2}}(0,v+h)^T\,\quad \Phi = v_\phi + \phi\,.
\end{equation}
The $\lambda_{H\phi}(H^\dagger H)\Phi^2$ term in the potential 
then leads to a non-zero mixing among the scalars and the 
squared mass matrix can be written as,
\bea
M_{h,\phi}^2 =
  \begin{pmatrix}
    2\l_H v^2 & \l_{H\phi} v v_{\phi} \\
    \l_{H\phi} v v_{\phi} &  2\l_{\phi} v_{\phi}^2 
  \end{pmatrix}\,.
\eea
The physical scalar masses $h_{1,2}$ are obtained upon 
diagonalizing the mass matrix, 
\bea
\begin{pmatrix}
    h_1 \\
    h_2
  \end{pmatrix} = \begin{pmatrix}
    c_\theta & -s_\theta \\
    s_\theta & c_\theta 
  \end{pmatrix}
  \begin{pmatrix}
    h \\
    \phi 
  \end{pmatrix}
\eea
with the mixing angle and mass eigenvalues given by
\begin{align}\label{eq:mixing}
    \tan{2\theta} &= \frac{ \l_{H\phi} v v_{\phi}}{\l_{\phi} v_{\phi}^2-\l_H v^2 }\,,\\
    M^2_{h_1,h_2} &= \big(\l_H v^2 +  \l_{\phi} v^2_{\phi}\big) \mp \sqrt{(\l_H v^2 -  \l_\phi v^2_{\phi}\big)^2 + \l_{H\phi}^2 v^2 v_{\phi}^2}\,.
\end{align}
Using the expressions above, we can rewrite the scalar 
potential couplings in terms of the masses and the 
mixing angle as
\begin{align}\label{dependent_couplings}
\l_{H} &= \frac{M_{h_1}^2c_{\theta}^2+M^2_{h_2}s_{\theta}^2}{2v^2}\,,\hspace{0.5em}
\l_{\phi} = \frac{M_{h_2}^2c_{\theta}^2+M^2_{h_1}s_{\theta}^2}{2v_{\phi}^2}\,,\hspace{0.5em}
\l_{H\phi}= \frac{(M_{h_2}^2 -M_{h_1}^2)c_{\theta}s_{\theta}}{vv_{\phi}}\,.   
\end{align}
Further, constraints from co-positivity require $\l_{H,\phi}>0,\l_{H\phi}<2\sqrt{\l_H \l_\phi}$, whereas perturbativity demands $\l_{H,\phi}\,,\l_{H\phi}\leq 4\pi$. These constraints are taken into account while doing the analysis.

In the limit, $v_\phi \gg v$ and $\lambda_{H\phi} \sim \lambda_\phi \sim \lambda_H$, we get $\tan{2\theta}\ll 1$ from Eq.\eqref{eq:mixing}, which corresponds to the small mixing 
limit with $c_\theta \sim 1$ and $s_\theta \ll 1$. In this 
limit $M_{h_1} \sim M_h$ and $M_{h_2} \sim M_\phi$, where we 
identify $h$ with the observed Higgs boson with mass 
125.25 GeV\cite{ParticleDataGroup:2022pth}. Hence, the 
phenomenology of our model will depend on the following 
independent parameters: $\{\L,M_\phi,M_N,M_{\chi},v_{\phi}\}$, 
where we adopt $M_\chi \ll M_{N,\phi} \ll \Lambda$. 

\section{Dark matter relic abundance}\label{sec:DM}

In the present setup, the role of DM is played by the fermion 
$\chi$ that acquires mass $M_\chi=y_{\chi}v_\phi$, once 
$\zf$ symmetry is spontaneously broken to a remnant 
$\mathcal{Z}_2$ when $\Phi$ acquires a non-zero VEV. 
This remnant $\mathcal{Z}_2$ symmetry also guarantees 
the stability of the DM.
A sufficiently large VEV suggests a very feeble coupling of TeV or a sub-TeV DM with the scalar $\Phi$. In such a scenario, the DM is produced with a negligible initial abundance that gradually increases over time. Depending on the mass of $\Phi$ the production of the DM from $\Phi$ decay can continue in the epoch between $\zf$ symmetry breaking and electroweak symmetry breaking i.e. in the regime $T^\ast > T > T'$, where $T^\ast(T')$ denotes the temperature of $\zf$ (electroweak) symmetry breaking. For $T < T'$, due to the mixing between the scalars, $\chi$ is produced in the decays of $h_{1,2}$, however, one of them would be suppressed due to the small value of mixing angle $\theta$\footnote{Other production channels via scattering such as ${\rm SM~SM}\r \chi \bar{\chi}$ or $N l \r \chi \bar{\chi}$ are further suppressed than the decays due to this small mixing.}. As a result of this feeble interaction, the DM particles never reach thermal equilibrium with the bath particles and their abundance freezes-in once the number density of the parent particle becomes Boltzmann suppressed. 

A sizeable mass for $\chi$ and a large $v_\phi$ ($y_\chi \ll 1$) thus ensure that the non-thermality condition ($\Gamma<H$) is always satisfied, where $\Gamma$ is the decay width given by
\begin{align}
   \Gamma_{\phi\to\chi\chi}=\frac{y_{\chi}^2}{8\pi}\,M_{\phi}\,\bigg(1-\frac{4M_{\chi}^2}{M_{\phi}}\bigg)^{\frac{3}{2}}=\frac{M_\phi}{8\pi}\,\left(\frac{M_\chi}{v_\phi}\right)^2\,\bigg(1-\frac{4M_{\chi}^2}{M_{\phi}^2}\bigg)^{\frac{3}{2}}\,,
\end{align}
for $T^\ast > T > T'$. Here, we have traded the feeble coupling $y_\chi$ in terms of our independent parameters $M_\chi$ and $v_\phi$. Similarly for $T < T'$, the decay width for the processes $h_{1,2} \r \chi \bar{\chi}$ can be written as 
\begin{align}
   \Gamma_{h_2\to\chi\chi}=\frac{M_{h_2} M_\chi^2 c_\theta^2}{8\pi v_\phi^2}\,\bigg(1-\frac{4M_{\chi}^2}{M_{h_2}^2}\bigg)^\frac{3}{2}\,,\quad
   \Gamma_{h_1\to\chi\chi}=\frac{M_{h_1} M_\chi^2 s_\theta^2}{8\pi v_\phi^2}\,\bigg(1-\frac{4M_{\chi}^2}{M_{h_1}^2}\bigg)^\frac{3}{2}\,.
\end{align}
We include the above processes for DM production by assuming a 
small value of the mixing angle ($s_\theta \sim 0.1$) for completeness, as for lighter values of DM mass, $\chi$ can also 
be produced from the decays of the SM Higgs boson. Apart, we can have 
scattering processes producing the DM as well, but they will be
suppressed compared to the decay contribution at temperatures close 
to DM mass. 

The evolution of the co-moving number density of the DM particle 
can be studied by solving the appropriate Boltzmann 
equation, however, in the case of decay dominated production, 
one can obtain the relic abundance by the following analytical expression \cite{Hall:2009bx}
\begin{equation}\label{eq:dmrelic}
    \Omega_{\chi}h^2\approx 1.09\times10^{27}\frac{g_\phi}{g_s \sqrt{g_{\rho}}}\,\left(\frac{M_\chi}{\text{GeV}}\right)\,\sum_X\frac{ 
  \Gamma_{X\rightarrow\chi\chi}}{M^{2}_{X}}\,,
\end{equation}
where $X = \phi$ for $T > T'$ and $X = h_1,h_2$ for $T < T'$, $g_\phi$ is the number of internal degrees of freedom associated 
with the scalar $\phi$, and $g_s$ and $g_\rho$ are the 
effective number of degrees of freedom in the bath associated 
with entropy and energy density at $T=M_\phi$ respectively. 
From Eq.~\eqref{eq:dmrelic}, it can be seen that the correct 
relic abundance can be produced for a suitable choice of 
$M_\phi$ and $v_\phi$ for a given DM mass $M_\chi$.
\begin{figure}[!htb]\label{fig:dmrelic}
\centering
\includegraphics[scale=0.5]{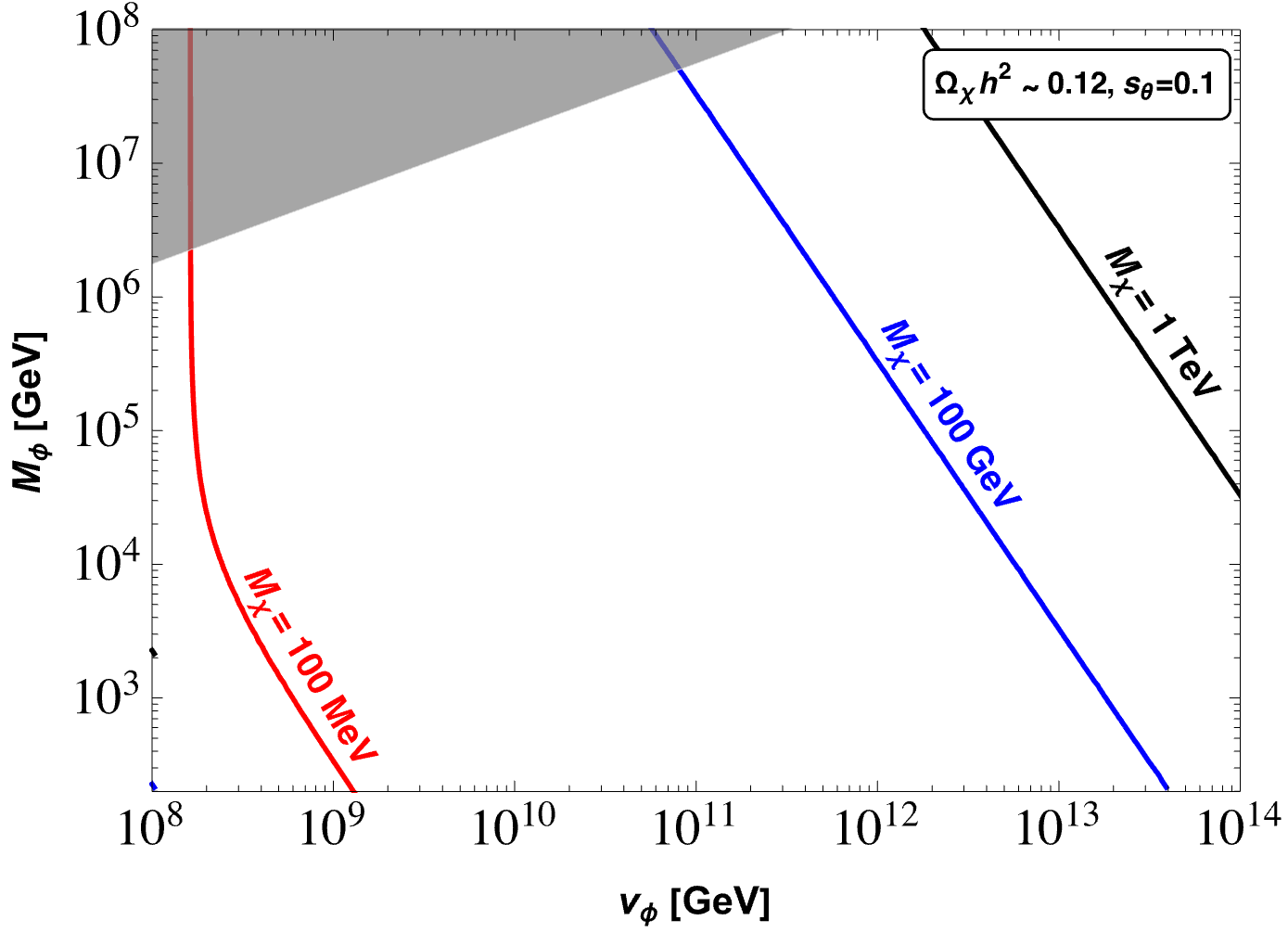}
\caption{Contours of $M_\chi$ that satisfy the observed DM relic abundance $\Omega_{\rm DM} h^2 = 0.12\pm 0.0012$ depending on the mass and VEV of the singlet scalar. The shaded portion on the top-right corner represents the region excluded by perturbativity constraints on the scalar potential parameter $\l_{H\phi}$ for $s_\theta \sim 0.1$.}
\end{figure}
In Fig.~\ref{fig:dmrelic}, we show the contours of $M_\chi$ 
that satisfy the correct relic abundance for a given value 
of $\{M_\phi, v_\phi\}$, thus the model can easily accommodate 
MeV-GeV scale dark matter. The behaviour of the relic 
contours observed in Fig.~\ref{fig:dmrelic} can be understood 
as follows. Two features are observed in the contour with DM mass 100 MeV: (i) if the $v_\phi$ is small and $\phi$ mass is large, the DM 
is dominantly produced from the Higgs decay and the production 
from $\phi$ remains negligible and hence the contour 
remains independent of $\phi$ mass, this can also be 
understood by looking at Eq.~\ref{eq:dmrelic}, (ii) the 
pattern changes for a lighter $M_\phi$ and a variation with 
$v_\phi$ is observed. This is because the production of the 
DM in this regime gets more or less equal contribution from the 
decay of both the scalars. For a heavier DM mass, the production 
only takes place from the decay of $\phi$ and hence one can see 
the variation of $M_\phi$ with $v_\phi$ in blue and black contours. 

\section{Neutrino masses}\label{sec:numass}

As discussed above, the scalar VEV $v_\phi$ also generates an effective Yukawa coupling for the RHNs which can thus produce the light neutrino masses via the type-I seesaw mechanism (see Eq.~\eqref{eq:numass}). In order to be consistent with neutrino oscillation data, we use Cassas-Ibarra (CI) parameterization \cite{Casas:2001sr} to write the effective coupling matrix as,
\begin{equation}\label{eq:ci}
    y_\nu =\frac{\sqrt{2}}{v}\,U_{\rm PMNS}^\ast\,\sqrt{m_\nu^d}\, \mathcal{R}^T\,\sqrt{M_R}\,.
\end{equation}
Here $m_\nu^d\,(M_R)$ is the diagonal light neutrino (RHN) mass matrix, and $U_{\rm PMNS}$ is the unitary matrix that diagonalises $m_\nu = U_{\rm PMNS}^\ast m_\nu^d U_{PMNS}$ and $\mathcal{R}$ is a $3 \times 3$ orthogonal matrix, which we choose as,
\begin{align}
    \mathcal{R}=\begin{pmatrix}
        0 & \cos{z_R} & \sin{z_R} \\
        0 & -\sin{z_R} & \cos{z_R}\\
        1 & 0 & 0
    \end{pmatrix}\,,
\end{align}
with a complex angle $z_R = a+ib$. For example, considering the normal hierarchy and taking the lightest active neutrino to be massless,
\begin{align}
   m_\nu^d = {\rm Diag}\{0,\sqrt{\Delta m_{21}^2},\sqrt{\Delta m_{32}^2+\Delta m_{21}^2}\}\nonumber\,,
\end{align}
where $\Delta m_{32}^2=2.511 \times 10^{-3}{\rm~eV^2}, \Delta m_{21}^2=7.41 \times 10^{-5} {\rm~eV^2}$ correspond to the mass squared difference from atmospheric and solar neutrino oscillations \cite{Esteban:2020cvm} with $\Delta m_{ij}^2 \equiv m_i^2 -m_j^2$; $M_R = {\rm Diag}\{5 \times 10^{11},10^{13},10^{17}\}$ and $z_R = 0.6367-0.115i$. This results in, 
\begin{equation}\label{eq:yukci}
    y_{\nu}=\begin{pmatrix}
        $ 0.00311326 - 0.000989169 i\ $ & $- 0.0283498 - 0.00877828 i\ $  & 0\\
 $0.0186216 - 0.00189537 i\ $ & $~0.0585416 + 0.0108465 i\ $ & 0\\
 $0.00585001 - 0.00259809 i\ $ & $~0.0876781 + 0.0041639 i\ $ &  0
\end{pmatrix}\,.
\end{equation}
In rewriting the coupling as $y_{\nu_{\alpha i}} \equiv h_{\alpha i} v_\phi/\Lambda$, the effect of $v_\phi$ cannot be seen directly, however a large enough value of $v_\phi$ is consistent with $\mathcal{O}(1)$ values for $h$ and high scale $\Lambda$ to generate a Yukawa matrix satisfying the oscillation data. It can be seen that opting for a FIMP like DM in this model allows for a wider and relatively unconstrained range of $v_\phi$. This can be contrasted with the WIMP DM considered in Ref.~\cite{Bhattacharya:2018ljs}, where the constraints on the DM mass (generated similarly via $v_\Phi$) from relic density and direct search yields $v_\phi$ to be $\mathcal{O}(10^2-10^3)$ GeV, thus implying stronger constraints on $h$ and $\Lambda$. 

\section{Leptogenesis}\label{sec:lepto}

When employing the type-I seesaw mechanism, the Majorana mass term for the RHNs breaks the lepton number and subsequently the out-of-equilibrium CP violating decays of the right-handed neutrinos $N \r l H$ can thus satisfy all the necessary conditions \cite{Sakharov:1967dj} to dynamically generate a net lepton asymmetry via the leptogenesis mechanism, which can be later processed into a baryon asymmetry via electroweak sphalerons. In the following, we discuss the CP asymmetry produced in such decays as well as the contribution of new scattering processes in the model.

\subsection{CP asymmetry from decays}

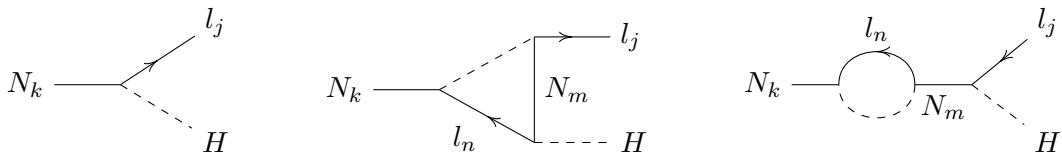
\begin{figure}[!htb]
    \centering
    \begin{tikzpicture}
    \begin{feynman}
    \vertex(i1) {$N_k$};
    \vertex[right=1.25 cm of i1] (i2);
    \vertex[right=1.25 cm of i2] (e);
    \vertex[above=0.5 cm of e] (a) {$l_j$};
    \vertex[below=0.5 cm of e](b) {$H$};
    \diagram* {
    (i1)-- (i2)--[fermion2] (a),
    (i2)-- [scalar](b),
    };
    \end{feynman}
    \end{tikzpicture}
    \hspace{2em}
    \begin{tikzpicture}
    \begin{feynman}
    \vertex(i1) {$N_k$};
    \vertex[right=1.25 cm of i1] (i2);
    \vertex[right=1.25 cm of i2] (e);
    \vertex[above=0.7 cm of e] (a);
    \vertex[below=0.7 cm of e](b);
    \vertex[right=1.0 cm of a](c) {$l_j$};
    \vertex[right=1.0 cm of b](d) {$H$};
    \diagram* {
    (i1)-- (i2)--[scalar] (a),
    (b)-- [fermion2, edge label=\(l_n\)](i2),
    (a)--[edge label =\(N_m\)] (b),
    (a)-- [fermion2](c),
    (b)-- [scalar](d)
    };
    \end{feynman}
    \end{tikzpicture}
    \hspace{2em}
    \begin{tikzpicture}
    \begin{feynman}
    \vertex(i1) {$N_k$};
    \vertex[right=1.0 cm of i1] (i2);
    \vertex[right=1.0 cm of i2] (f); 
    \vertex[right=0.75 cm of f] (g);
    \vertex[right=1.0 cm of g] (e);
    \vertex[above=0.5 cm of e] (a) {$l_j$};
    \vertex[below=0.5 cm of e](b) {$H$};
    \diagram* {
    (i1)-- (i2),(f)--[fermion2,half right,edge label'=\(l_n\)] (i2),(f) --[scalar, half left] (i2),
    (f)--[edge label' =\(N_m\)] (g),
    (a)-- [fermion2] (g)-- [scalar] (b),
    };
    \end{feynman}
    \end{tikzpicture}
    \caption{Tree level and one loop diagrams for the decay $N_k \rightarrow l_j H$ that contribute to the CP asymmetry parameter $\varepsilon_D$.}
    \label{fig:cpdecay}
\end{figure}

Due to the complex nature of Yukawa couplings $y_\nu$, a CP asymmetry is generated in $N_i$ decays via the interference between the tree and loop level decay amplitudes shown in Fig.~\ref{fig:cpdecay}, and can be written as
\begin{align}
    {\varepsilon_D}_i \equiv \frac{\Gamma(N_i\rightarrow l  H)-\Gamma(N_i \rightarrow \bar{l} \bar{H})}{\Gamma(N_i\rightarrow l  H)+\Gamma(N_i \rightarrow \bar{l} \bar{H})} \,,
\end{align}
where $\Gamma$ is the decay width of the RHN given by
\begin{equation}\label{eq:decayrhn}
    \Gamma(N_1\rightarrow lH)=\frac{(y_\nu^\dagger~y_\nu)}{16\pi}M_1\,.
\end{equation}
The lepton asymmetry is then generated if these decays happen out of equilibrium. Considering a hierarchical spectrum of RHNs, the dominant contribution to the lepton asymmetry comes from the lightest RHN, $N_1$ as any asymmetry produced by $N_2$ and $N_3$ decays is rapidly washed out by $N_1$ interactions at later times.
An explicit calculation of the interference term gives \cite{Covi:1996wh}
\begin{align}
    {\varepsilon_D}_1 = \frac{1}{8\pi} \sum_{m \neq 1} \frac{\text{Im}[(y_\nu^\dagger y_\nu)^2_{1m}]}{(y_\nu^\dagger y_\nu)_{11}}\bigg\{f_v \bigg(\frac{M_m^2}{M_1^2}\bigg) + f_s\bigg(\frac{M_m^2}{M_1^2}\bigg) \bigg\}\,,
\end{align}
where
\begin{align}\label{eq:vsfdecay}
    f_v(x)=\sqrt{x}\bigg[1-(1+x)\ln{\bigg(\frac{1+x}{x} \bigg)} \bigg];\quad
    f_s(x)=\frac{\sqrt{x}}{1-x}\,,
\end{align}
where $f_{v,s}$ represent the vertex and self-energy  corrections respectively. The final form of the CP asymmetry due to decays can then be written as
\begin{align}\label{eq:cpdecayapp}
   {\varepsilon_D}_1 =\frac{3}{16\pi}\, \sum_{m \neq 1}\frac{1}{\sqrt{x}}\, \frac{\text{Im}[(y_{\nu}^\dagger y_{\nu})^2_{1m}]}{(y_{\nu}^\dagger y_{\nu})_{11}}\,{\rm for}~x\gg 1\,,
\end{align}
with $x={M_m^2}/{M_1^2}$.

\subsection{CP asymmetry from scatterings}

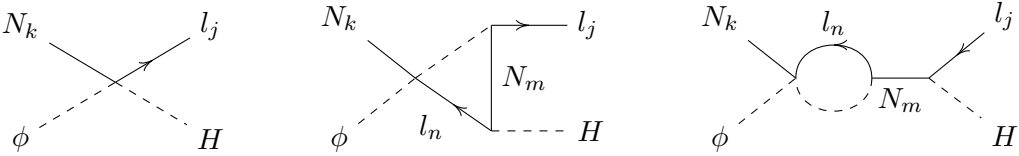
\begin{figure}[!htb]
    \centering
    \begin{tikzpicture}
    \begin{feynman}
    \vertex(i1) {$N_k$};
    \vertex[right=2.5cm of i1](i2) {$l_j$};
    \vertex[below=1.5 cm of i1](a) {$\phi$};
    \vertex[below=1.5 cm of i2](b) {$H$};
    \vertex[below=0.75cm of i1](c);
    \vertex[right=1.25cm of c] (e);    
    \diagram* {
    (i1)-- (e)--[fermion2] (i2),
    (a)-- [scalar](e) --[scalar] (b),
    };
    \end{feynman}
    \end{tikzpicture}
    \hspace{2em}
    \begin{tikzpicture}
    \begin{feynman}
    \vertex(i1);
    \vertex[above=0.5cm of i1] (g) {$N_k$};
    \vertex[below=0.5cm of i1] (h) {$\phi$};
    \vertex[right=1.0 cm of i1] (i2);
    \vertex[right=1.0 cm of i2] (e);
    \vertex[above=0.7 cm of e] (a);
    \vertex[below=0.7 cm of e](b);
    \vertex[right=1.0 cm of a](c) {$l_j$};
    \vertex[right=1.0 cm of b](d) {$H$};
    \diagram* {
    (g)-- (i2)--[scalar] (a),
    (h)--[scalar] (i2),(b)-- [fermion2, edge label=\(l_n\)](i2),
    (a)--[edge label =\(N_m\)] (b),
    (a)-- [fermion2](c),
    (b)-- [scalar](d)
    };
    \end{feynman}
    \end{tikzpicture} 
    \hspace{2em}
    \begin{tikzpicture}
    \begin{feynman}
    \vertex(i1);
    \vertex[above=0.5cm of i1] (k) {$N_k$};
    \vertex[below=0.5cm of i1] (h) {$\phi$};
    \vertex[right=1.0 cm of i1] (i2);
    \vertex[right=1.0 cm of i2] (f); 
    \vertex[right=0.75 cm of f] (g);
    \vertex[right=1.0 cm of g] (e);
    \vertex[above=0.5 cm of e] (a) {$l_j$};
    \vertex[below=0.5 cm of e](b) {$H$};
    \diagram* {
   (f)--[fermion2,half right,edge label'=\(l_n\)](i2) ,(f) --[scalar, half left] (i2), (h)--[scalar](i2), (k) --(i2),
    (f)--[edge label' =\(N_m\)] (g),
    (a)-- [fermion2] (g)-- [scalar] (b),
    };
    \end{feynman}
    \end{tikzpicture} 
    \caption{Tree and one-loop scattering diagrams for $N\phi \r l H$ that contribute to the CP asymmetry parameter $\varepsilon_S$.}
     \label{fig:cpscat}
\end{figure}

In the present setup, a CP asymmetry can also be generated via the scatterings of the form $N \phi \r l H\,(\bar{l} \bar{H})$ via the interference between the tree and one-loop amplitudes as shown in Fig.~\ref{fig:cpscat}. 

First, we define the reaction density for a $2 \leftrightarrow 2$ scattering process as
\begin{align}\label{eq:reactscat}
      \gamma(ij\rightarrow12) 
      & =\frac{T}{512 \pi^6}\int \text{d}\tilde{s}\,\frac{|\textbf{p}||\textbf{q}|}{\sqrt{\tilde{s}}}\,K_1\left(\frac{\sqrt{\tilde{s}}}{T}\right)\int d\Omega\, |\mathcal{M}|^2_{ij\rightarrow12}\,, 
\end{align}
where T denotes the temperature, $|\mathcal{M}|^2$ is the amplitude squared matrix element (summed over initial and averaged over final states), $K_1$ denotes the order one modified Bessel's function of the second kind, $\bf{p}(\bf{q})$ is the initial (final) state momentum in the center of mass frame and $\Omega$ is the solid angle. The integration variable $\tilde{s}$ runs from $\tilde{s}_{min}=\text{max}\big[(m_i+m_j)^2,(m_1+m_2)^2\big]$ (lower limit)  to $\tilde{s}_{max}=\Lambda$ (upper limit). The CP asymmetry produced due to scatterings $N\phi \r lH,\,\bar{l}\bar{H}$ can then be parameterized as
\begin{align}\label{eq:scattform}
    {\varepsilon_S}_i \equiv \frac{\gamma(N_i
    \phi\rightarrow l  \bar{H})-\gamma(N_i\phi \rightarrow \bar{l} H)}{\gamma(N_i\phi\rightarrow l  \bar{H})+\gamma(N_i\phi \rightarrow \bar{l} H)} \,.
\end{align}

Using Eq.~\eqref{eq:scattform}, we can express the CP asymmetry produced in the scatterings as \cite{Goudelis:2022bls}
\begin{align}\label{eq:cpasymform}
    {\varepsilon_S}_i\approx\, & \frac{\int\text{d}\tilde{s}\,\left({|\textbf{p}||\textbf{q}|\,}/{\sqrt{\tilde{s}}}\right)\,K_1\left({\sqrt{\tilde{s}}}/{T}\right)\,\int\left\{\left|\mathcal{M}\right|^2_{N_i\phi\rightarrow l  H}\,-\,\left|\mathcal{M}\right|^2_{N_i\phi \rightarrow \bar{l} \bar{H}}\right\}d\Omega}{2\int\text{d}\tilde{s}\,\left({|\textbf{p}||\textbf{q}|\,}/{\sqrt{\tilde{s}}}\right)\,K_1\left({\sqrt{\tilde{s}}}/{T}\right)\,\int\left|\mathcal{M}\right|^2_{N_i\phi\rightarrow l  H}\big|_0 d\Omega}\nonumber \\
    =&-2\,\frac{\text{Im}\big\{c_0^*c_1\big\}}{|c_0|^2}\frac{\int\text{d}\tilde{s}\,\left({|\textbf{p}||\textbf{q}|\,}/{\sqrt{\tilde{s}}}\right)\,K_1\left({\sqrt{\tilde{s}}}/{T}\right)\,\int\text{Im}\big\{\mathcal{A}_0^*\mathcal{A}_1\big\}d\Omega}{\int\text{d}\tilde{s}\,\left({|\textbf{p}||\textbf{q}|\,}/{\sqrt{\tilde{s}}}\right)\,K_1\left({\sqrt{\tilde{s}}}/{T}\right)\,\int\left|\mathcal{A}_0\right|^2d\Omega}\,,
 \end{align}   
where we have expressed the matrix element as a product of the coupling constant $c$ and the amplitude $\mathcal{A}$, so that $ \mathcal{M}|_j\equiv \sum^j_{e=0}c_e\mathcal{A}_e$, with $j$ signifying the loop order (here $j=0,1$). 

\begin{figure}[!htb]
\centering
\includegraphics[scale=0.5]{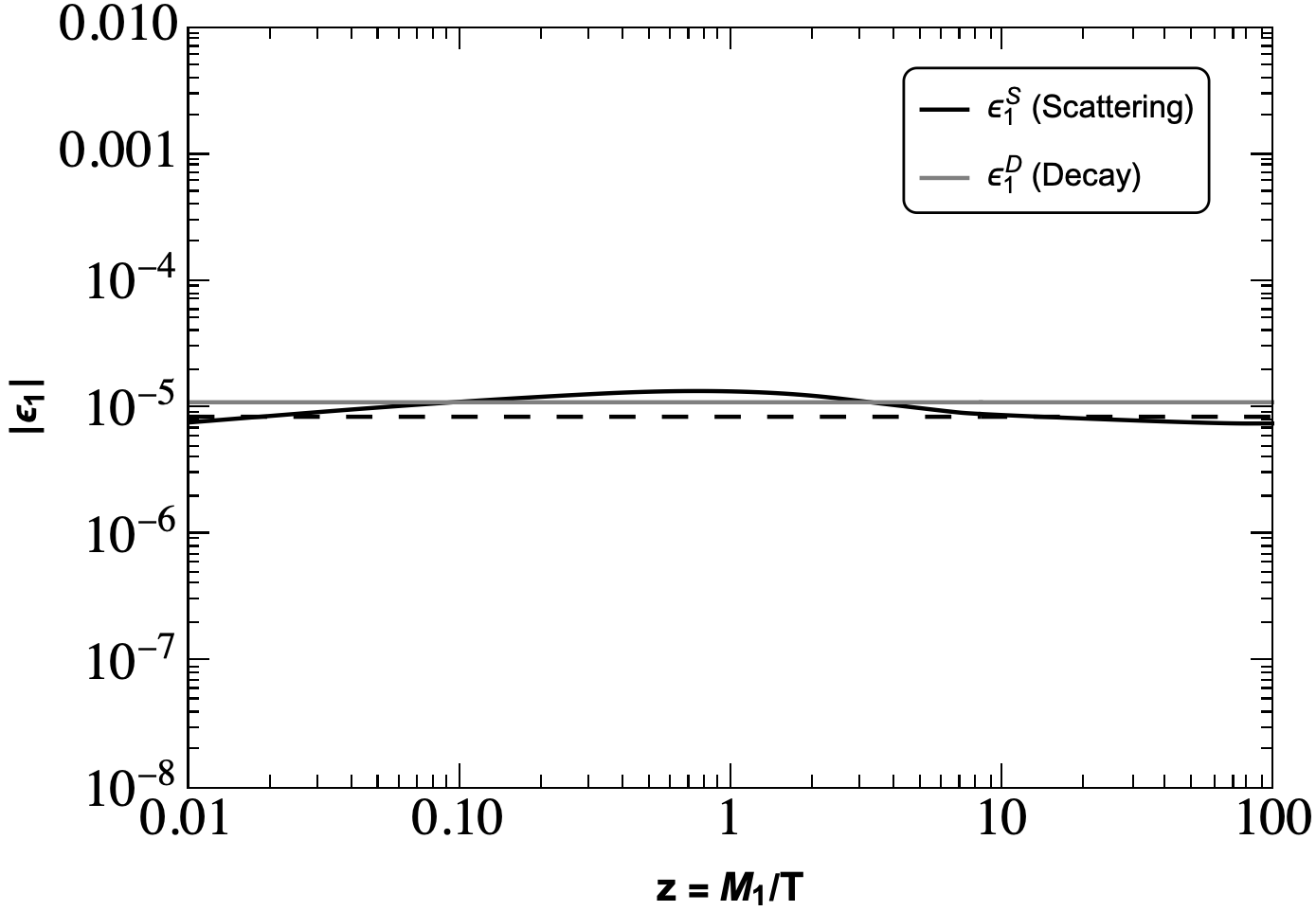}
\caption{The evolution of $\varepsilon_1$ with $z \equiv M_1/T$ due to the scattering and decay processes, taking $M_1 = 5 \times 10^{11}{\rm~GeV}\,,M_2 = 10^{13}{\rm~GeV}\,,M_\phi = 10^8{\rm~GeV}$ and $a=0.6367,b=0.115$. The black solid (dashed) curve represents the numerical (analytical) solution for ${\varepsilon_S}_1$ while the solid gray represents the CP asymmetry produced only via the decays for the same choice of parameters.}
\label{fig:decayscatasym}
\end{figure}

Note that the numerator of Eq.~\eqref{eq:cpasymform} is proportional to the interference between the one loop-level and tree-level amplitudes with ${\rm Im}\{c_0^\ast c_1\} = {\rm Im}[(y_\nu^\dagger y_\nu)^2_{im}]/v_\phi^{2}$ and
\begin{align}    
     {\rm Im}\big\{\mathcal{A}_0^*\mathcal{A}_1\big\}= -\frac{M^2_i}{32 \pi}\sqrt{x}\,\left\{1-(1+x') \ln\left[\frac{1+x'}{x'}\right]\right\}-\frac{\tilde{s}}{64 \pi}\left(\frac{\sqrt{x}}{1-x}\right)\,,
\end{align}
where $x = M_m^2/M_i^2$ and $x' = M_m^2/\tilde{s}$. The above expressions can be contrasted with the expressions for the vertex and self-energy corrections in Eq.~\eqref{eq:vsfdecay} for the case of decays, with the notable difference being the inclusion of the parameter $\tilde{s}$ denoting c.o.m. energy squared. On the other hand, the denominator of Eq.~\eqref{eq:cpasymform} is proportional to the total tree level scattering reaction density,
\begin{equation}\label{eq:gammas}
    \gamma(N_i \phi \r l H)=\frac{T}{512\pi^6} \frac{{(y_\nu^\dagger y_\nu)}_{ii}}{v_\phi^2}\int\text{d}\tilde{s}\,\frac{{\sqrt{\lambda(\tilde{s},M^2_\phi,M^2_i)}/4}}{\sqrt{\tilde{s}}}\,K_1\left(\frac{\sqrt{\tilde{s}}}{T}\right)(\tilde{s}+M^2_i-M^2_\phi)\pi\,,
\end{equation}
where we have used $|c_0|^2 = {{(y_\nu^\dagger y_\nu)}_{ii}}/{v_\phi^2}$ and $\left|\mathcal{A}_0\right|^2={(M^2_i-M^2_\phi+\tilde{s})}/{4}$. Also, here $\lambda$ is the Källén function. The integral needs to be evaluated numerically, however, in the high energy limit for $\tilde{s} \gg M_{i,m}$, the above expression can be approximated as,
\begin{equation}\label{eq:reactapprox}
    \gamma(N_i \phi \r l H)\approx \frac{{(y_\nu^\dagger y_\nu)}_{ii}}{v_\phi^2}\frac{T^6}{64\pi^5} =\frac{(y_\nu^\dagger~y_\nu)}{64\pi^2}\frac{M_1^4}{z}f_c(M_1,v_\phi,z)\,,
\end{equation}
where $f_c(M_1,v_\phi,z)={M_1^2}/({v_\phi^2}{\pi^3 z^5})$ is a dimensionless quantity. In writing the above equations, we have factored the function $f_c$ so that it can be contrasted with the decay rate,
\begin{equation}\label{eq:reactiondensitydecay}
    \gamma(N_i\rightarrow lH)=\frac{(y_\nu^\dagger~y_\nu)}{64\pi^2}\frac{M_1^4}{z}K_1(z)\,,
\end{equation}
with $z\equiv M_{1}/T$. Similar to the case of decays, for a hierarchical spectrum of RHN masses, the dominant contribution to CP asymmetry from scatterings will come from $N_1$ scatterings only ($i=1$)
\begin{align}
        {\varepsilon_S}_1=&-2\sum_{m \neq 1} \frac{\text{Im}[(y_{\nu}^\dagger y_{\nu})^2_{1m}]}{(y_{\nu}^\dagger y_{\nu})_{11}}\frac{\int\text{d}\tilde{s}\,{\sqrt{\lambda(\tilde{s},~M^2_\phi,~M^2_1)}}/{(4\sqrt{\tilde{s}})}\,K_1\left({\sqrt{\tilde{s}}}/{T}\right)\,\int\text{Im}\big\{\mathcal{A}_0^*\mathcal{A}_1\big\}d\Omega}{\int\text{d}\tilde{s}\,{\sqrt{\lambda(\tilde{s},~M^2_\phi,~M^2_1)}}/{(4\sqrt{\tilde{s}})}\,K_1\left({\sqrt{\tilde{s}}}/{T}\right)\,\int\left|\mathcal{A}_0\right|^2d\Omega}\,.
\end{align}
In deriving the formula above, we have traded the $h/\Lambda$ coupling by $y_\nu/v_\phi$, notice that the dependence on $v_\phi$ cancels. Substituting the expressions for $\mathcal{A}$ above, we can make an analytical approximation in the high energy limit similar to Eq.~\eqref{eq:reactapprox},
\begin{equation}
    {\varepsilon_S}_1 \approx \frac{2}{16\pi}\sum_{m \neq 1}\frac{1}{\sqrt{x}} \frac{\text{Im}[(y_{\nu}^\dagger y_{\nu})^2_{1m}]}{(y_{\nu}^\dagger y_{\nu})_{11}}\quad ({\rm for~} \tilde{s}\gg M_{i,m}{\rm~and~}x\gg1)\,.
\end{equation}
The approximation is strikingly similar to the expression we obtained for decays (see Eq.~\eqref{eq:cpdecayapp}), as expected given the similarity between the decay and scattering processes in the set-up. In Fig,~\ref{fig:decayscatasym}, we compare the contribution to CP asymmetry from scatterings ($N\phi \r lH$) and decays ($N\r lH$) corresponding to the same set of benchmark values.   

We also compare the approximate result with the numerical solutions and find that they are in good agreement for lower values and higher values of $z$, though, there is a slight deviation around $z\sim 1$ since $\tilde{s} \sim M_1$ and there is an enhancement from the terms that we neglect while making the approximations. When evaluating the lepton asymmetry via the relevant Boltzmann equations as we show below, we use both the numerical and the analytical results and find that the approximation works well for the range of values we are interested in.

It is worth pointing out that the scattering contribution discussed in our model is qualitatively different from the existing works where such contributions have been taken into account for generating the asymmetry. For example, in Refs.~\cite{Bento:2001rc, Gu:2009yx,Baldes:2014gca, Baldes:2014rda,Ghosh:2021ivn}, the focus was to generate baryon asymmetry predominantly from $2 \leftrightarrow 2$ scatterings and the decay contribution is either sub-dominant or rendered ineffective. In the case of WIMPy Baryogenesis, the baryon asymmetry is generated from the DM annihilations responsible for freeze-out~\cite{Cui:2011ab, Bernal:2012gv, Bernal:2013bga, Kumar:2013uca, Baldes:2015lka,Racker:2014uga}. On the contrary, here both the decay and scattering processes generate a CP asymmetry, while the scattering processes contribute mainly to wash-out, as discussed below. Further, the scatterings in the present set-up are a consequence of the $\mathcal{Z}_4$ symmetry invariant effective interaction of the model and do not involve any dark sector particles.

\subsection{Boltzmann equations}

To study the evolution of lepton asymmetry of the universe, we have to solve the following set of coupled Boltzmann equations (BEQs) for the number densities of $\phi$, $N$ and $\Delta L = n_l - n_{\bar{l}}$\footnote{As we are only interested in the dyanamics of lightest RHN, to simplify the notation we drop the subscript $1$, and denote $N_1,{\varepsilon_D}_1$ and ${\varepsilon_S}_1$ as simply $N,\varepsilon_D$ and $\varepsilon_S$, unless stated otherwise.} 
\begin{align}
    \frac{dY_{\phi}}{dz}&=-\frac{1}{sHz}\left[\gamma_S \left(\frac{Y_{N}Y_{\phi}}{Y_{N}^{eq}Y_{\phi}^{eq}}-1\right)+\gamma_{\phi hh}\left(\frac{Y_{\phi}}{Y_{\phi}^{eq}}-1\right)+s^2 \langle\sigma v\rangle_{\phi H} \left(Y_{\phi}^2-{Y_{\phi}^{eq}}^2\right)\right]\,,\label{eq:beqphi}\\
     \frac{dY_{N}}{dz}&=-\frac{1}{sHz}\left[\gamma_D\left(\frac{Y_{N}}{Y_{N}^{eq}}-1\right)+\gamma_S\left(\frac{Y_{N}Y_{\phi}}{Y_{N}^{eq}Y_{\phi}^{eq}}-1\right)\right]\,,\label{eq:beqn}\\
    \frac{dY_{\Delta L}}{dz}&= \frac{1}{sHz}\left[\gamma_D \varepsilon_D\left(\frac{Y_{N}}{Y_{N}^{eq}}-1\right)+\gamma_S \varepsilon_S\left(\frac{Y_{N}Y_{\phi}}{Y_{N}^{eq}Y_{\phi}^{eq}}-1\right)-\frac{Y_{\Delta L}}{2Y_L^{eq}}(\gamma_D+\gamma_S)\right]\,,\label{eq:beql}
\end{align}
where $Y^{eq}=n^{eq}/s$, $s=0.44 g_s T^3$, $H=1.66 g_\rho^{1/2} T^2/M_{Pl}$, $z\equiv M_{1}/T$ and $\gamma$ denotes the reaction density. It is defined in Eq.~\eqref{eq:reactscat} for scatterings, whereas in case of decays, $\gamma_{abc}=n^{eq}_{a}\,\Gamma(a\rightarrow bc)\,{K_1(z)}/{K_2(z)}$. 

Further, $\gamma_D$ is the total $N$ decay width $\gamma(N\rightarrow LH)+\gamma(N\rightarrow \Bar{L}\Bar{H})$ (see Eq.~\eqref{eq:decayrhn}) and $\gamma_S$ is the total $N$ scattering cross section $\gamma_S\equiv\gamma(N\phi\rightarrow LH)+\gamma(N \phi\rightarrow \Bar{L}\Bar{H})$, see Eq.~\eqref{eq:gammas}. In addition to the above, we use the following expressions for the interactions of $\phi$ and $H$
\begin{equation}\label{eq:phiscatdecay}
    \Gamma(\phi\rightarrow hh)=\frac{\lambda_{H\phi}^2}{16\pi~M_\phi} v_\phi^2\,,\quad
    \langle\sigma v\rangle_{\phi H}=\frac{\lambda_{H\phi} }{64\pi M_\phi^2}\,\left(1-\frac{M_h^2}{M_\phi^2} \right)^{1/2}\,,
\end{equation} 
where $\langle\sigma v\rangle_{\phi H}$ is the thermal average crosssection for the process $\phi\phi\rightarrow HH$~\cite{McDonald:1993ex}. Now, we discuss the terms originating from various processes in the BEQs one by one.

In writing the BEQ for $\phi$, we have neglected its decays to the DM $\chi$ as its contribution will be sub-dominant compared to other terms, given the feeble nature of coupling and also omitted the BEQ for $\chi$, as an analytical solution has already been provided in Section~\ref{sec:DM}. The first term in Eq.~\eqref{eq:beqphi} comes from scatterings with $N$, whereas the second and third term correspond to the decay and scattering with the SM Higgs via the terms in Eq.~\eqref{eq:scalarpot} for the scalar potential. To ensure that $\phi$ remains in thermal equilibrium with the SM bath, we impose the constraint $\Gamma(\phi \r hh)>H$ at $T=M_\phi$, since we expect the decays to two Higgs to dominate over the scatterings, and obtain the following bound
\begin{equation}\label{eq:lambdabound}
    \lambda_{H\phi} > 0.04 \left(\frac{10^8{\rm~GeV}}{v_\phi}\right)\,\left(\frac{M_\phi}{10^{10}{\rm~GeV}}\right)^{3/2}\,.
\end{equation}
Note that $\lambda_{H\phi}$ also depends on the mixing angle $\theta$ along with $M_\phi$ and $v_\phi$, as can be seen from Eq.~\eqref{dependent_couplings}. 

In Eqs.~\eqref{eq:beqn} and \eqref{eq:beql} for the evolution of $N$ abundance and lepton asymmetry, we have included the scattering contribution analogous to the decays in type-I leptogenesis. However, in order to derive the correct BEQ, it is essential to include all the processes (upto the leading order in the involved couplings), therefore, in addition to the decays $N \r l H (\bar{l} \bar{H})$, scatterings $N\phi \r l H (\bar{l} \bar{H})$ and their respective inverse processes\footnote{We do not consider the $\Delta L =1$ scatterings, since for the value of parameters that we work with, the washout from such prcoesses can be neglected.}, we also need to include the $\Delta L =2$ scatterings $l H \leftrightarrow \bar{l}{H}$ in order to obtain the correct sign accompanying $1$ in the brackets in the first and second term of Eq.~\eqref{eq:beql} which quantifies the deviation from thermal equilibrium, since the contribution from the on-shell part corresponding to $N$ and $N, \phi$ (via loop) mediated scatterings is already taken into account when considering the forward and backward reactions, and therefore must be subtracted in order to avoid double counting \cite{Strumia:2006qk}. Moreover, this can also be seen as a consequence of CPT and unitarity, $\sum_j |\mathcal{M}(i \r j)|^2 = \sum_j |\mathcal{M}(j \r i)|^2$, which gives
\begin{eqnarray}
    \sigma(l H \r \bar{l}\bar{H}) - \sigma(\bar{l}\bar{H} \r l H) = \epsilon_D \gamma_D + \epsilon_S\gamma_S \,.
\end{eqnarray}

\subsection{Numerical results}

To obtain the asymptotic value of lepton asymmetry ($Y_{\Delta L}^\infty$) produced via the decays and scatterings in our model, we solve the set of coupled Boltzmann equations numerically for a set of benchmark points (BPs) listed in Table~\ref{tab:bp}. In order to understand the role of the scalar singlet in leptogenesis, we are interested in investigating the following three scenarios: \textit{i)} $v_\phi \gg M_\phi$, \textit{ii)} $v_\phi \sim M_\phi$ and \textit{iii)} $v_\phi < M_\phi$, which is reflected in our choice of benchmark values.
\begin{table}[!htb]\label{tab:bp}
\centering
\begin{tabular}{c c c c c}
\hline \hline 
Benchmarks & $M_\phi$  (GeV)& $v_\phi$ (GeV)&  $z_R$  &$M_\chi$ \\
 \hline 
 BP1 & $1\times10^{7}$ & $1.5\times10^{12}$ &  $0.6367-0.00515i$ & $43$~\text{GeV},~~$400$~\text{GeV}\\
 BP2 &$1.4\times10^{10}$ & $1.4\times10^{10}$ &  $0.6367-0.0084i$ & $200$~\text{GeV}\\
  BP3 & $5\times10^9$ & $2.5\times10^{9}$ & $0.6367-0.056i$ &$45$~\text{GeV}\\
 \hline \hline
\end{tabular}
\caption{Set of benchmark values of $M_\phi$ and $v_{\phi}$ along with the corresponding value of $z_R$ that satisfy the observed baryon asymmetry as well as the corresponding DM mass that reproduces the observed DM relic abundance. Here, the lighter DM can also be produced from the SM Higgs due to the small mixing.}
\end{table}
For all the benchmark points, we fix the mass of the RHNs: $M_1 = 5 \times 10^{11}$ GeV and $M_2= 10^{13}{\rm~GeV},M_3= 10^{17}{\rm~GeV}$, so that the Yukawa coupling $y_\nu$ for each BP differs only due to the choice of $z_R = a +ib$ that enters the orthogonal matrix $\mathcal{R}$ in the CI parameterization of Eq.~\eqref{eq:ci}. The values of $a, b$ can then be tweaked to obtain the value of CP asymmetries ($\varepsilon_{D,S}$) that produced the correct order of lepton asymmetry, which in turn reproduces the observed baryon asymmetry of the universe, $Y_B = C_{sp} Y_{\Delta L}^\infty = (8.75 \pm 0.23) \times 10^{-11}$ \cite{ParticleDataGroup:2022pth}, where $C_{sp} =28/51$ is the sphaleron conversion factor. Further, we assume initial thermal equilibrium for the heavy RHNs. 

In the last column of Table~\ref{tab:bp}, we indicate the DM mass corresponding to the choice of $M_\phi$ and $v_\phi$ which reproduces the correct relic abundance. For \textbf{BP1}, it can be seen from Eq.~\eqref{dependent_couplings} that it is possible to have a sizeable mixing between $\phi$ and $H$, i.e., $s_\theta \sim 0.1$. Therefore, in this case, the DM may also be produced in the SM Higgs decays via this small mixing, the lighter DM mass (43 GeV) corresponds to this contribution. For the other two BPs, this mixing is highly suppressed and the DM is produced only from $\phi$ decays.  

\begin{figure}[!htb]
    \centering
    \includegraphics[width=0.48\linewidth]{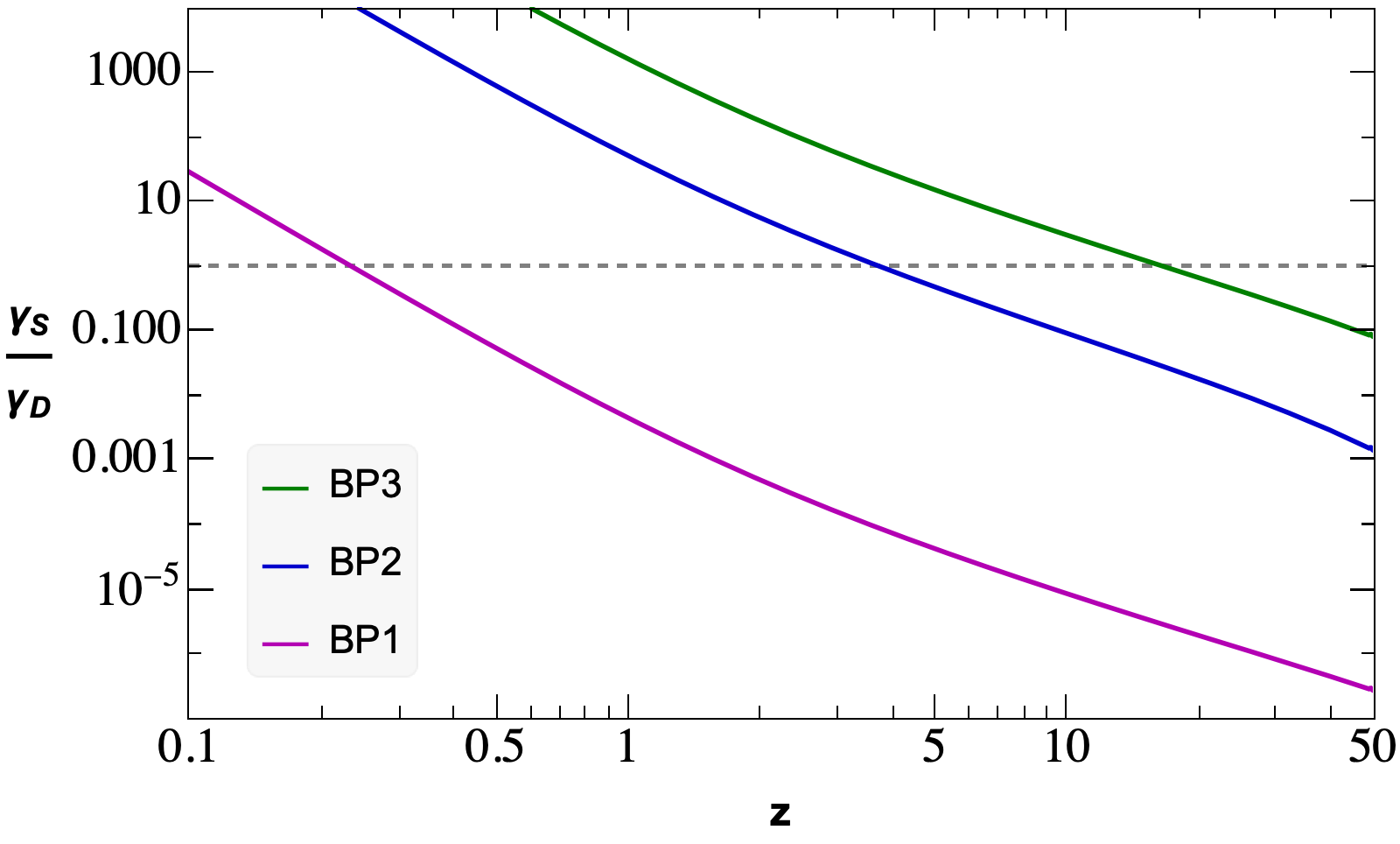}
    \includegraphics[width=0.48\linewidth]{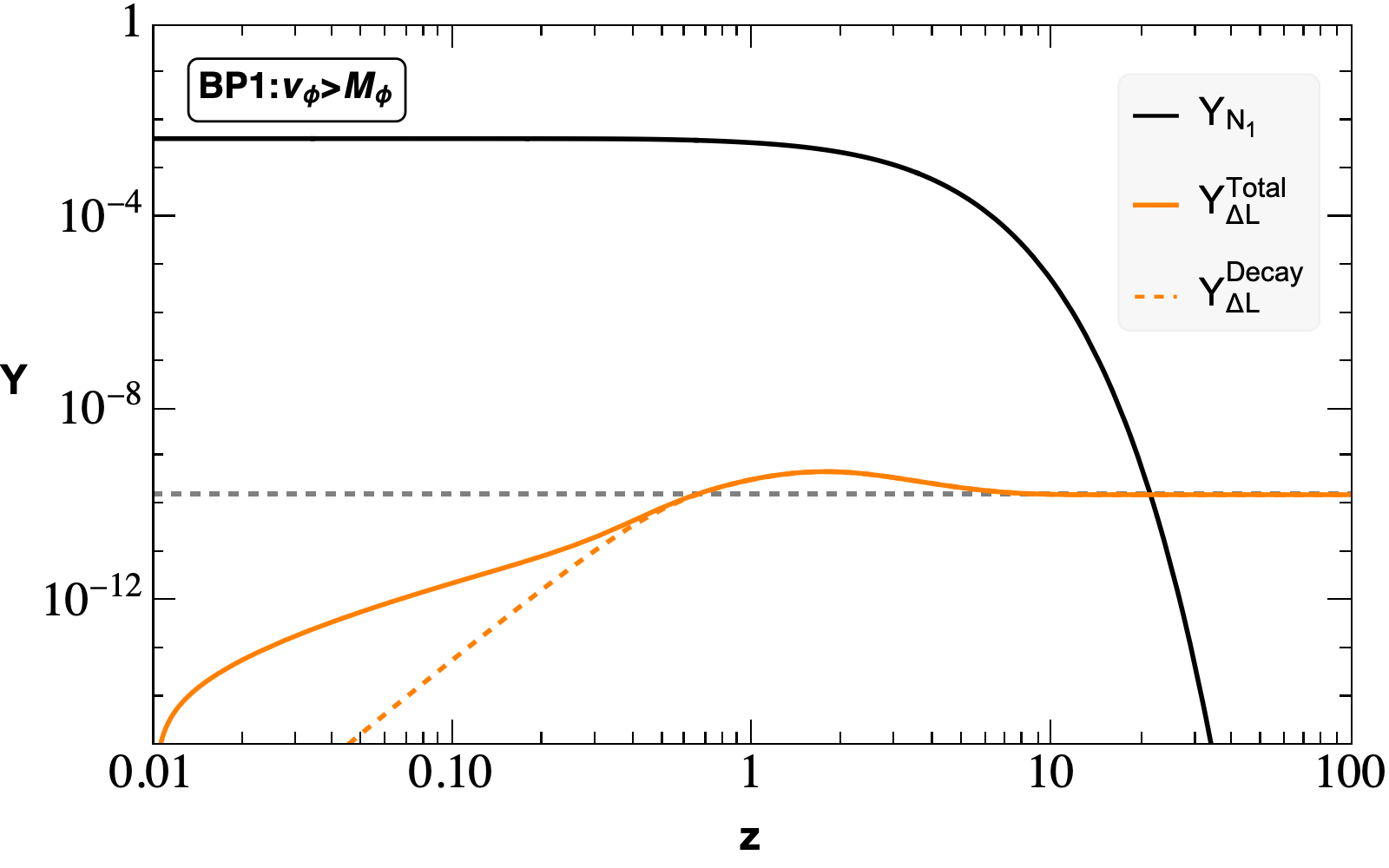}
    \caption{\emph{Left:} Comparison of scattering and decay rates for our choice of benchmark points. The dashed line indicates that they are equal. \emph{Right:} Numerical solutions of the Boltzmann equations corresponding to {\bf BP1}. The evolution of $Y_{N_1}$ ($Y_{\Delta_L}$) is depicted by the solid black (orange) curve. The orange dashed curve denotes the evolution of lepton asymmetry in the absence of scatterings, whereas as the gray dashed line indicated the correct $Y_{\Delta L}$ which satisfies the observed baryon asymmetry of the universe.}
    \label{fig:BEQsol12}
\end{figure}

We compare the reaction rate of scatterings to that of decays for our three BPs in the left plot of Fig.~\ref{fig:BEQsol12}. Since $\gamma_S \propto v_\phi^{-2}$, it can be seen that the scatterings are suppressed than decays for our choice of {\bf BP1} corresponding to $v_\phi \gg M_\phi$, whereas they are stronger than decays until $z \sim 10$ for the other two choices, where we expect them to play an important role in determining the asymptotic lepton asymmetry. 

In the right plot of Fig.~\ref{fig:BEQsol12}, we plot the numerical solutions for {\bf BP1}. The black curve denotes the evolution of $N_1$ abundance starting from thermal equilibrium. On the other hand, the lepton asymmetry (denoted by the orange solid curve) sourced by the first and second term in Eq.~\eqref{eq:beql} starts rising from its zero initial value until $z \sim 1$, whereas the third term leads to a partial washout of this asymmetry via the inverse process. Once the temperature drops below $M_1$, these processes are disfavored by kinematics and the lepton asymmetry thus freezes-in around $z \sim 10$. We also plot the evolution of lepton asymmetry in the absence of scatterings $N\phi \r lH (\bar{l}\bar{H})$, i.e, $Y_{\Delta L}$ is sourced only via decays (orange dashed curve). It is worth pointing that the $N_1$ abundance curve deviates slightly from its equilibrium distribution (necessary for generation of a lepton asymmetry), however, the deviation is quite small to be visible in the plots.

It can be seen that the total lepton asymmetry is initially much larger than the asymmetry produced just by decays indicating that the scatterings source the asymmetry for $z\ll 1$. The dynamics can be understood as follows, for large values of $v_\phi$, the scattering reaction density is small, therefore, the interactions of $\phi$ with $H$ keep it in equilibrium. This leads to an enhancement from the second term in Eq.~\eqref{eq:beql} for $z \ll 1$. Although, the scattering dominates initially, for $z>1$, the solid and dashed orange curves coincide indicating that the decay contribution dominates over the scatterings (see the purple curve in Fig.~\ref{fig:BEQsol12}) and the asymptotic lepton asymmetry can be obtained just from decays. Hence, we find that in the high $v_\phi$ regime, the scattering contributions can be neglected and we recover the canonical type-I leptogenesis mechanism.
\begin{figure}[!htb]
\centering
\includegraphics[width=0.48\linewidth]{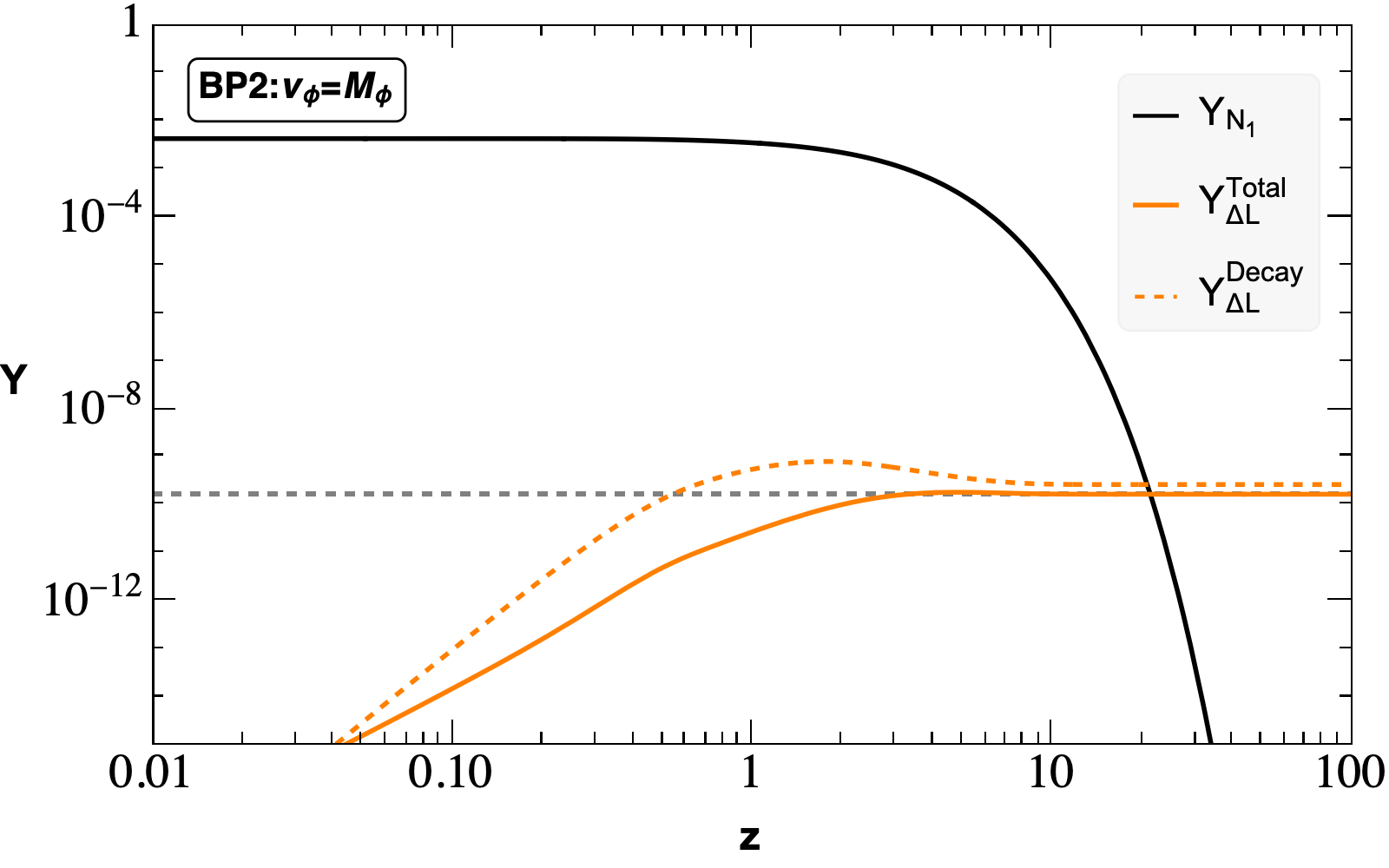}
\includegraphics[width=0.48\linewidth]{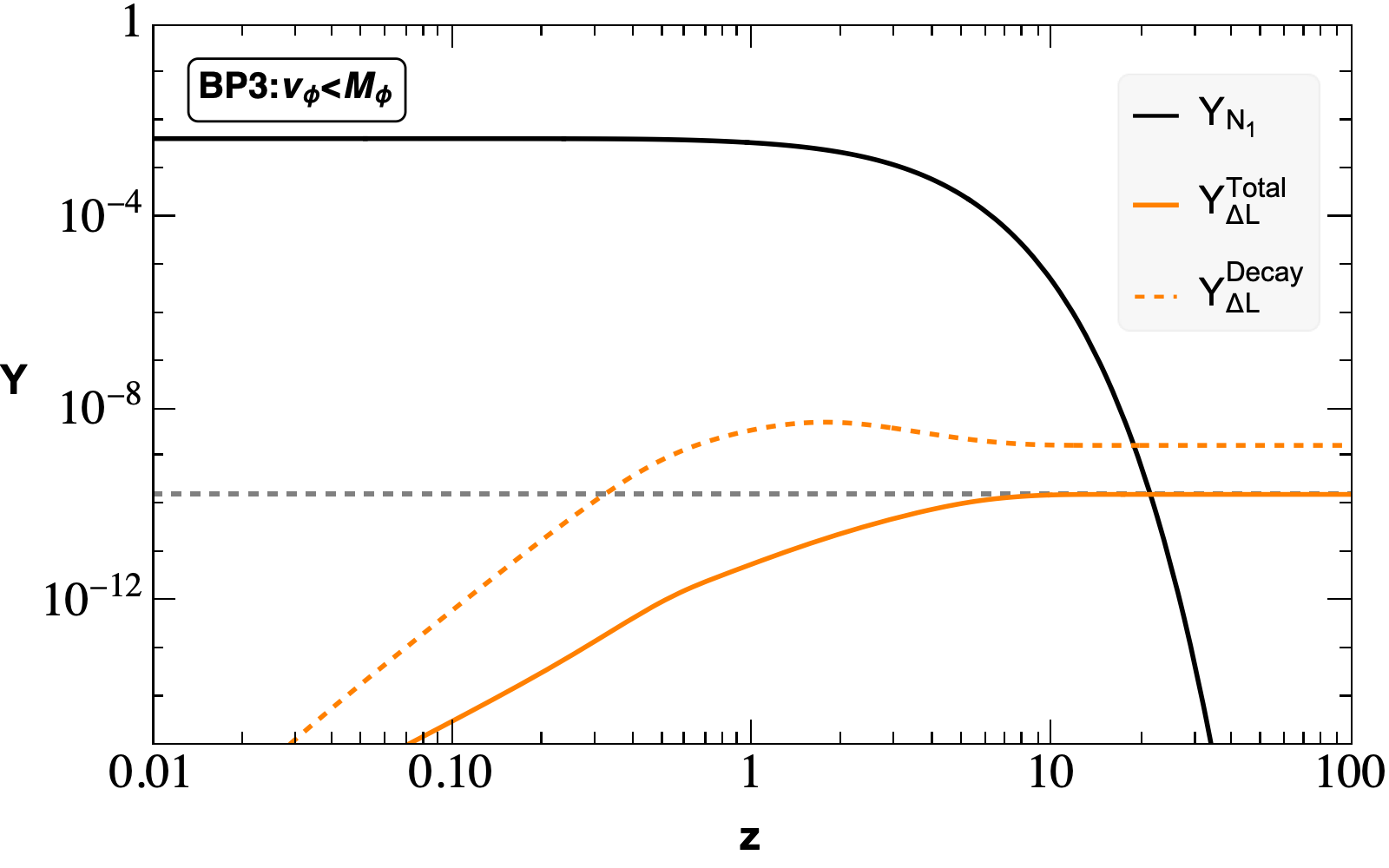}
\caption{Numerical solutions of the Boltzmann equations for {\bf BP2} and {\bf BP3}, similar to the right plot of Fig.~\ref{fig:BEQsol12}.}\label{fig:BEQsol1}
\end{figure}

Unlike the solution for {\bf BP1}, the total lepton asymmetry evolves differently initially for the case of {\bf BP2} and {\bf BP3}, as shown in Fig.~\ref{fig:BEQsol1}. In the left plot, as $v_\phi$ is comparable to $M_\phi$, the scattering contribution is larger than before. Notice that the total lepton asymmetry is slightly smaller than the one that would have been generated solely in decays, indicating a mild washout of the decay generated asymmetry via the scatterings. Hence, in this regime, the scattering contributions must be taken into account to reproduce the required value to lepton asymmetry to match observations.

Finally, for the case where $v_\phi < M_\phi$, as seen in the right plot of Fig.~\ref{fig:BEQsol1}, the total lepton asymmetry is much smaller than the one that would have been generated just in decays, when compared to {\bf BP2}. This is attributed to the fact that the scattering reaction densities are quite large, which in turn leads to larger deviation from equilibrium for both $N_1$ and $\phi$ and a stronger washout of the decay generated asymmetry. Hence, when $v_\phi$ is smaller compared to $M_\phi$, we find that the presence of scatterings plays a larger role in the washout of the generated asymmetry than sourcing it and thus have a significant effect in leptogenesis for lower $v_\phi$ values. This implies, that for a fixed RHN mass and $v_\phi$, it must be bounded from below, i.e. $v_\phi > v_\phi^c$ in order to avoid too strong washout. For values less than $v_\phi^c$, one cannot reproduce the lepton asymmetry required to match the observations, and thus we can constrain the parameter space of our model from leptogenesis. For $M_1 = 5 \times 10^{11}$ GeV, we find that $v_\phi^c \sim 6 \times 10^{8}$ GeV.

\section{Gravitational wave signals}\label{sec:GW}

In this section, we discuss the possibility of testing our high-scale model. As discussed earlier, our setup remains invariant under a $\mathcal{Z}_4$ discrete symmetry until it is spontaneously broken to a remnant $\mathcal{Z}_2$ symmetry as a result of $\Phi$ getting a non-zero a VEV, $\langle\Phi\rangle=\pm v_\phi$. Unlike the DM $\chi$ all the particles including both SM and BSM remain even under the remnant $\mathcal{Z}_2$. The scalar $\Phi$ can take one of the two values, i.e., $+v_\phi~\text{and}-v_\phi$, resulting in the formation of two different domains and production of domain walls (DWs) from their boundaries \cite{Saikawa:2017hiv}. In principle, these DWs can be very long-lived and can dominate the energy budget of the Universe at some stage (their energy density falls much more slowly than that of matter and radiation, $i.e.~\rho_\text{DW}\propto a^{-1}$ with $a$ being the scale factor). This in turn modifies the evolution of the Universe in a way that is inconsistent with the current CMB observations. 

A possible solution to the DW problem is to introduce an energy bias in the potential that can lift the degenerate minima~\cite{Vilenkin:1981zs, Gelmini:1988sf, Larsson:1996sp}. This makes the DWs unstable and helps them collapse before overclosing the Universe. In the present setup, this can be achieved by introducing terms in the scalar potential that can softly break the discrete $\mathcal{Z}_4$. The simplest possibility that one can think of is,
\begin{equation}
    \Delta V=\mu\Phi^3+\mu_{\phi H}\Phi H^\dagger H+\frac{c_1}{\Lambda}\Phi^5+\frac{c_2}{\Lambda}\Phi^3H^\dagger H+\frac{c_3}{\Lambda}\Phi (H^\dagger H)^2,
    \label{deltaV}
\end{equation}
where $\mu$ and $\mu_{\phi H}$ have a mass dimension. Next, we also assume the coefficients $c_1,c_2$ and $c_3$ to be sufficiently small such that the contribution of all the dim-5 operators in Eq.~\eqref{deltaV} can safely be ignored\footnote{In principle, the dimension 5 operator  $\phi^5/\Lambda$  can also dominate   $ V_{\text{bias}}$ but in such a scenario, one requires a very large value of $\Lambda$ as was discussed in~\cite{King:2023ayw,King:2023ztb} to generate GW that can be observed by the present or future GW detectors. A relatively smaller $\Lambda$ will generate a very large $ V_{\text{bias}}$ which in turn will generate GW spectrum with a very large peak frequency which might be out of the reach of present and future detectors. On the other hand, one should also note that a DW might not be created if there exists a very large $ V_{\text{bias}}$ according to the prediction of percolation theory~\cite{Saikawa:2017hiv}. }. Once $\Phi$ obtains a VEV, the second term in Eq.\eqref{deltaV} can contribute to the mass of Higgs and hence we demand this contribution to be small or negligible.  Following Eq.~\eqref{deltaV}, the energy bias term can be expressed as
\begin{equation}
    V_{\text{bias}}=\mu v_\phi^3+\mu_{\phi H}\frac{v_\phi v^2}{2}.
    \label{Vbias}
\end{equation}
 Moreover, one also notices that in the limit  $v_\phi \gg v$ (as required by the dark matter, neutrino masses and baryon asymmetry in this unified framework ), the first term dominates $ V_{\text{bias}}$ until and unless $\mu$ is negligibly small and hence, the contribution of the second term in  Eq.~\eqref{Vbias} can be safely ignored. So, without any loss of generality, we set $\mu_{\phi H}=0$ for the rest of our analysis. 

 Once the degeneracy of the vacuum is uplifted one also demands that the population of the true vacuum should be greater than that of the false vacuum (one with the higher energy)~\cite{Gelmini:1988sf}. This results in the generation of volume pressure force $p_V\sim V_{\text{bias}}$ that acts on the walls, forcing the region of false vacuum to shrink. Once $p_V$ becomes greater than the tension force $p_T$ of the wall, the DWs start to collapse and annihilate. This produces a significant amount of gravitational waves (GWs) which may remain as a stochastic background in the present Universe. Under the assumption that the DWs annihilate in a radiation-dominated era and the annihilation happens instantaneously at $t=t_{\text{ann}}$, the peak frequency $f_{p}$ and peak energy density spectrum $\Omega_{\text{GW}}h^2$ of GWs at present can be expressed as
 \begin{align}\label{fpeak}
     f_{p}&\simeq 1.4\times10^{-5}~\text{Hz}\times\bigg(\frac{1.41}{\mathcal{A}}\bigg)^{1/2}\bigg(\frac{10^7~\text{GeV}}{\mathcal{\sigma}^{1/3}}\bigg)^{3/2}\bigg(\frac{V_{\text{bias}}}{10^7~\text{GeV}^4}\bigg)^{1/2}\,,\\
     \Omega_{p}h^2&\simeq 1.49\times10^{-10}\times\bigg(\frac{\tilde{\epsilon}_{\text{GW}}}{0.7}\bigg)\bigg(\frac{\mathcal{A}}{1.41}\bigg)^4\bigg(\frac{\sigma^{1/3}}{10^7~\text{GeV}}\bigg)^{12}\bigg(\frac{10^7~\text{GeV}^4}{V_{\text{bias}}}\bigg)^{2},
     \label{omegapeak}
 \end{align}
 where the efficiency factor $\tilde{\epsilon}_{\text{GW}}\simeq0.7$~\cite{Hiramatsu:2013qaa} can be regarded as a constant in the scaling regime and the area parameter is chosen as $\mathcal{A}=1.41\pm0.13$ following the axion model with $N=4$~\cite{Kawasaki:2014sqa}. Finally, $\sigma=\frac{2}{3}\sqrt{2\lambda_\phi}v_\phi^3$ is the surface energy density (surface tension) of the wall. To depict the GW spectrum, we adopt the following parametrization for a broken power-law spectrum~\cite{Caprini:2019egz, NANOGrav:2023hvm}
\begin{eqnarray}
 \Omega_{\rm GW}h^2_{} = \Omega_p h^2 \frac{(a+b)^c}{\left(b x^{-a / c}+a x^{b / c}\right)^c} \ ,
\label{eq:spec-par}
\end{eqnarray}
where $x \equiv f/f_p$, and $a$, $b$ and $c$ are real and positive parameters. Here the low-frequency slope $a = 3$ can be fixed by causality, while numerical simulations suggest $b \simeq c \simeq 1$~\cite{Hiramatsu:2013qaa}.
\begin{figure}[!htb]
\centering
\includegraphics[width=0.48\linewidth]{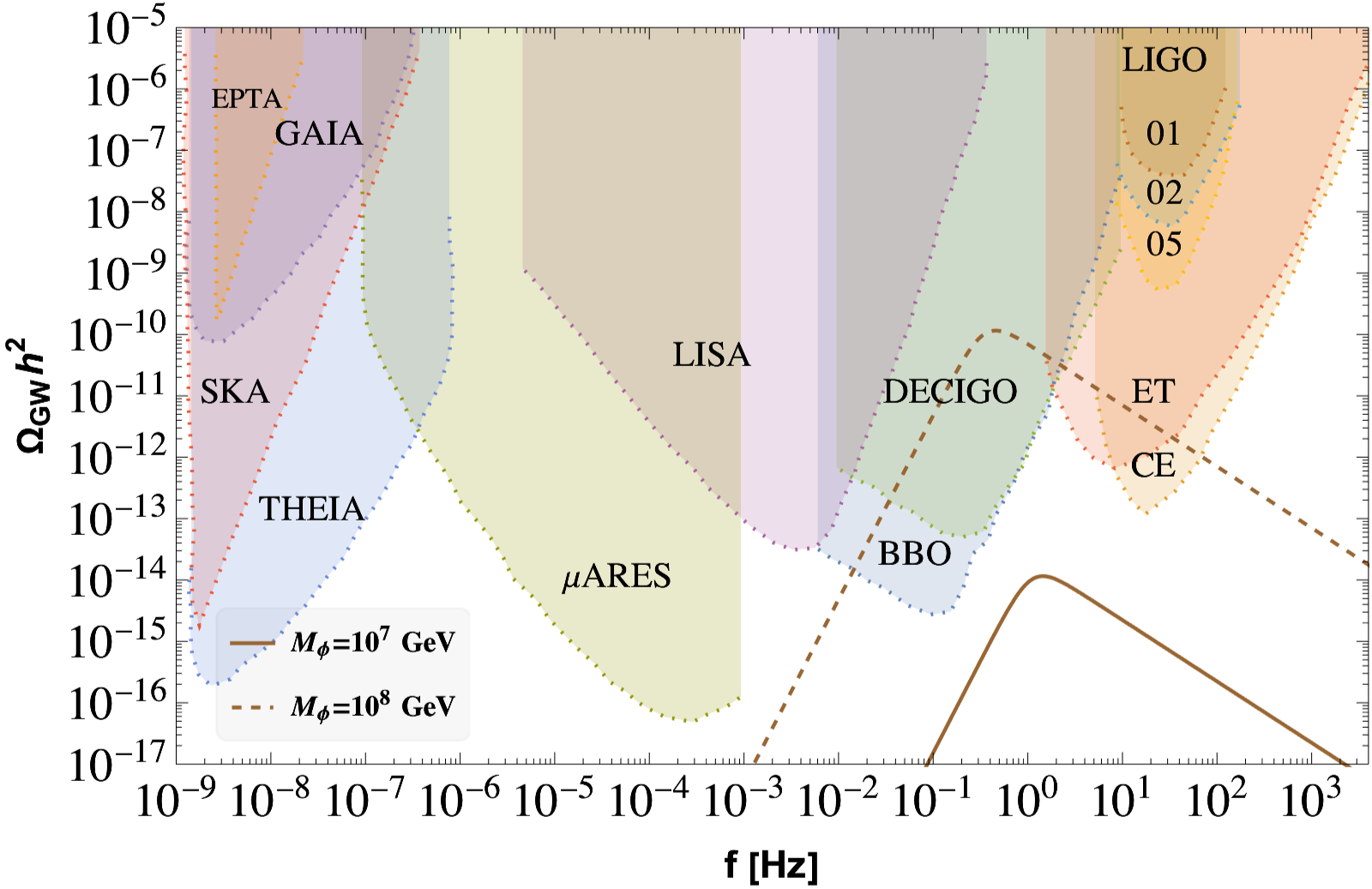}
\includegraphics[width=0.48\linewidth]{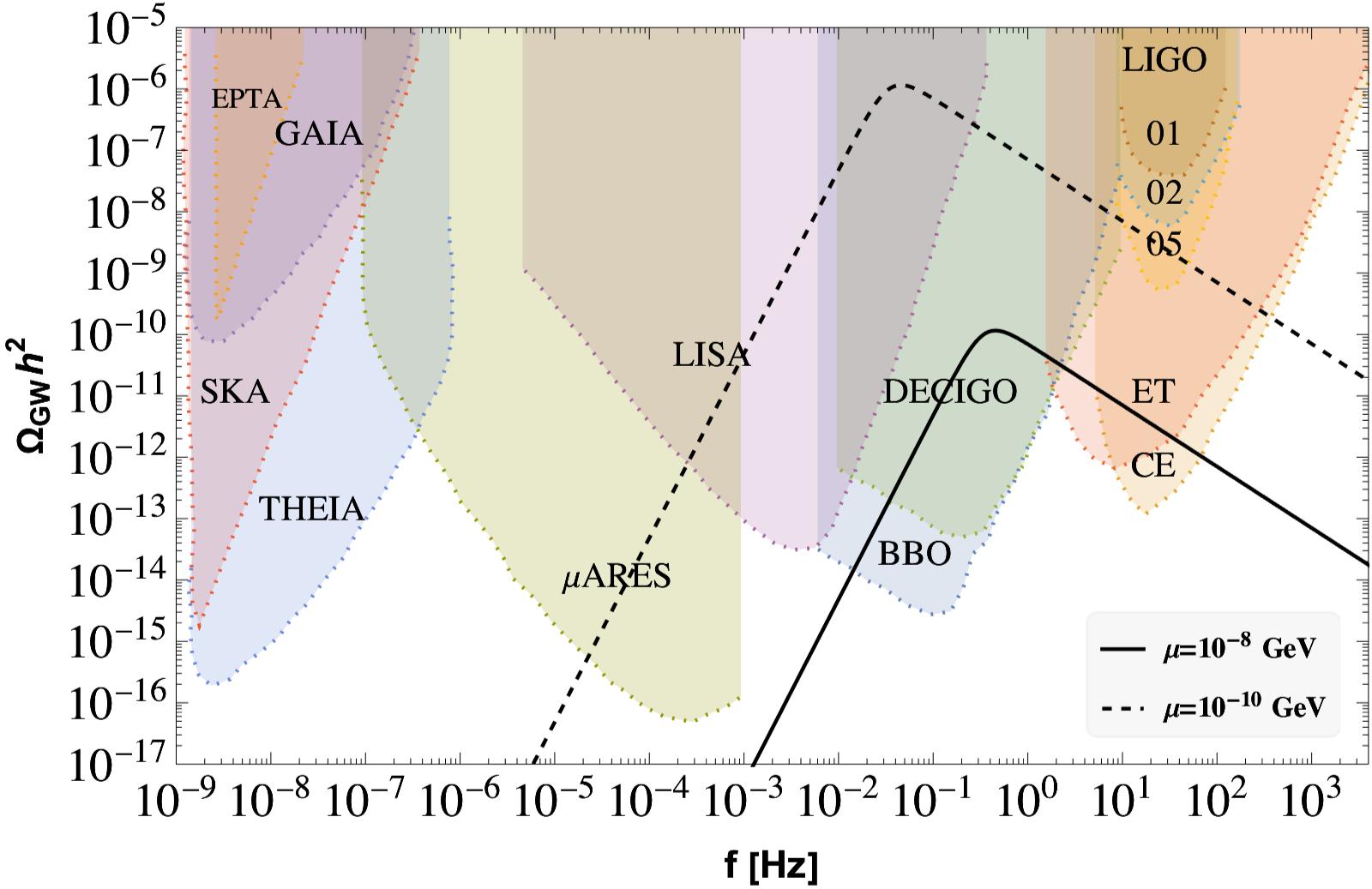}
\includegraphics[width=0.48\linewidth]{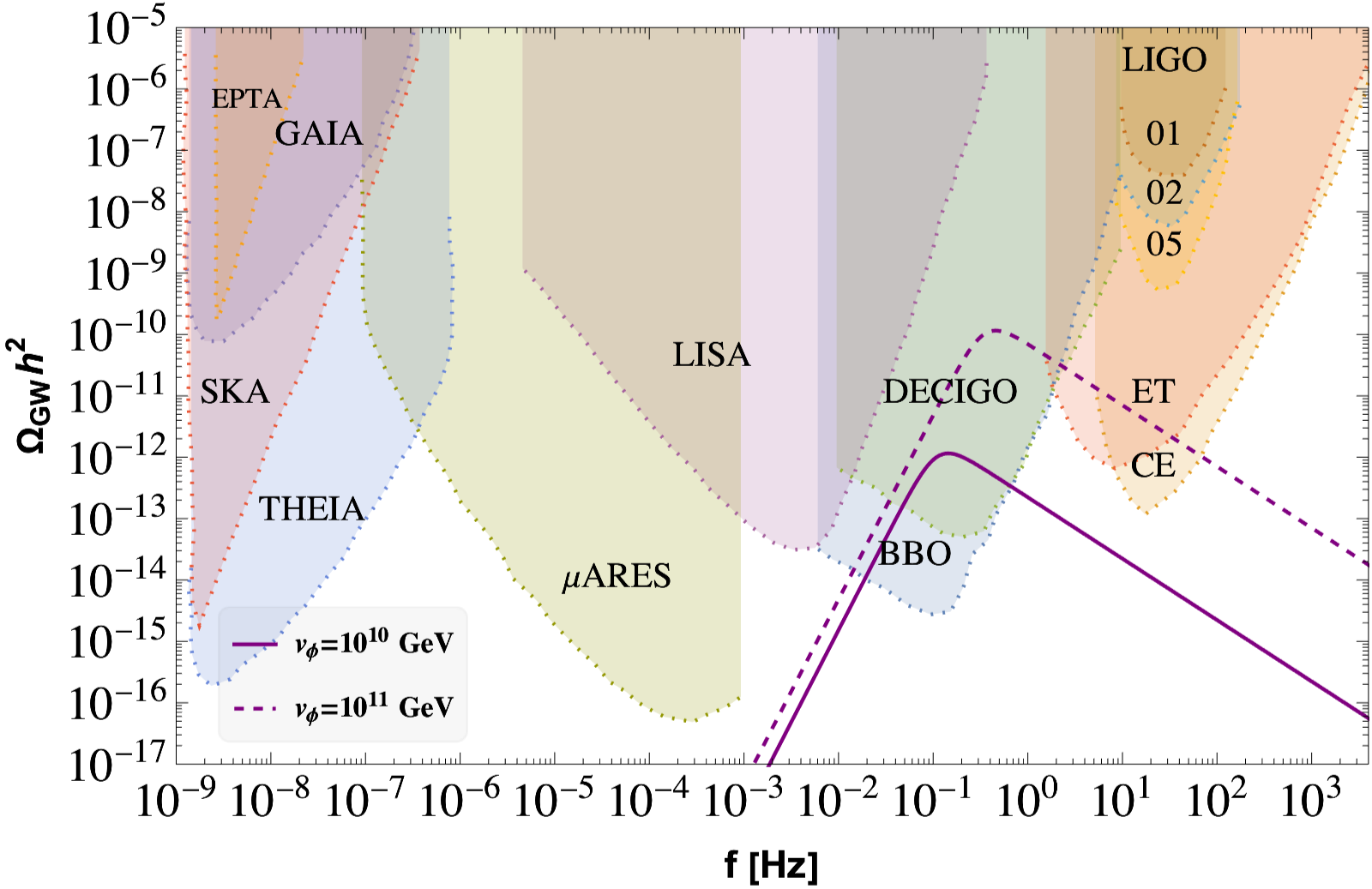}
\caption{Top left panel: $v_\phi=10^{11}$ GeV, $\mu=10^{-8}$ GeV; Top right panel: $v_\phi=10^{11}$ GeV, $M_{\phi}=10^{8}$ GeV; Bottom panel: $M_{\phi}=10^{8}$ GeV, $\mu=10^{-8}$ GeV }
\label{GW_spectrum}
\end{figure}
The corresponding GW spectrums are shown in Fig.~\ref{GW_spectrum} for different values of model parameters. In all of these plots, the experimental sensitivities of SKA~\cite{Weltman:2018zrl}, GAIA~\cite{Garcia-Bellido:2021zgu}, EPTA~\cite{Moore:2014eua}, THEIA~\cite{Garcia-Bellido:2021zgu}, $\mu$ARES~\cite{Sesana:2019vho}, LISA~\cite{LISA:2017pwj}, DECIGO~\cite{Seto:2001qf, Kawamura:2006up, Yagi:2011wg}, BBO~\cite{Crowder:2005nr, Corbin:2005ny, Harry:2006fi}, ET~\cite{Punturo:2010zz, Hild:2010id, Sathyaprakash:2012jk, Maggiore:2019uih}, CE~\cite{LIGOScientific:2016wof, Reitze:2019iox}, and aLIGO~\cite{LIGOScientific:2016wof, LIGOScientific:2014qfs, LIGOScientific:2016jlg} are shown as shaded regions of different colors. In the top left panel, for the demonstration purpose, we keep the model parameters $v_\phi= 10^{11}$ GeV and $\mu=10^{-8}$ GeV fixed while we vary the mass of $M_\phi$\footnote{Given that the $\mu$ term in Eq.~\eqref{deltaV} is a super-renormalizable interaction, the choice $\mu =10^{-8}$ GeV corresponds to an extremely tiny explicit $\mathcal{Z}_4$ symmetry breaking in the model. For the values of $M_\phi$ that we work with, $\frac{\mu}{M_\phi}\sim 10^{-15}-10^{-18}$, which is way smaller than the freeze-in coupling $y_\chi \sim 10^{-10} - 10^{-8}$ required for DM relic density.}. A larger $M_\phi$ corresponds to a larger value of $\lambda_\phi$ and hence a larger surface tension $\sigma$ is obtained. This in turn generates a large $\Omega_p h^2$ following Eq.~\ref{omegapeak} as observed in this plot. In the top right panel, we fix $M_\phi=10^8$ Gev while we keep $v_\phi$ at the same value. Varying $\mu$ in this plot directly affects $V_\text{bias}$ and hence a larger $\Omega_p h^2$ is observed for a smaller $\mu$ and vice-versa. Finally, in the bottom panel, we study the effect of varying $v_\phi$ on the GW spectrum while keeping $M_\phi=10^8$ GeV and  $\mu=10^{-8}$ GeV fixed. Varying $v_\phi$ affects both surface tension as well as the  $V_\text{bias}$ but $\Omega_p h^2$ is still dominated by $\sigma$ as it remains proportional to $\sigma^{4}$. Hence, a larger $v_\phi$ corresponds to a larger $\Omega_p h^2$. Depending on the different combinations of $\{M_\phi,v_\phi,\mu\}$ the GW spectrum remains within the reach of present and future GW experiments.

\begin{figure}[!htb]
\centering
\includegraphics[scale=0.5]{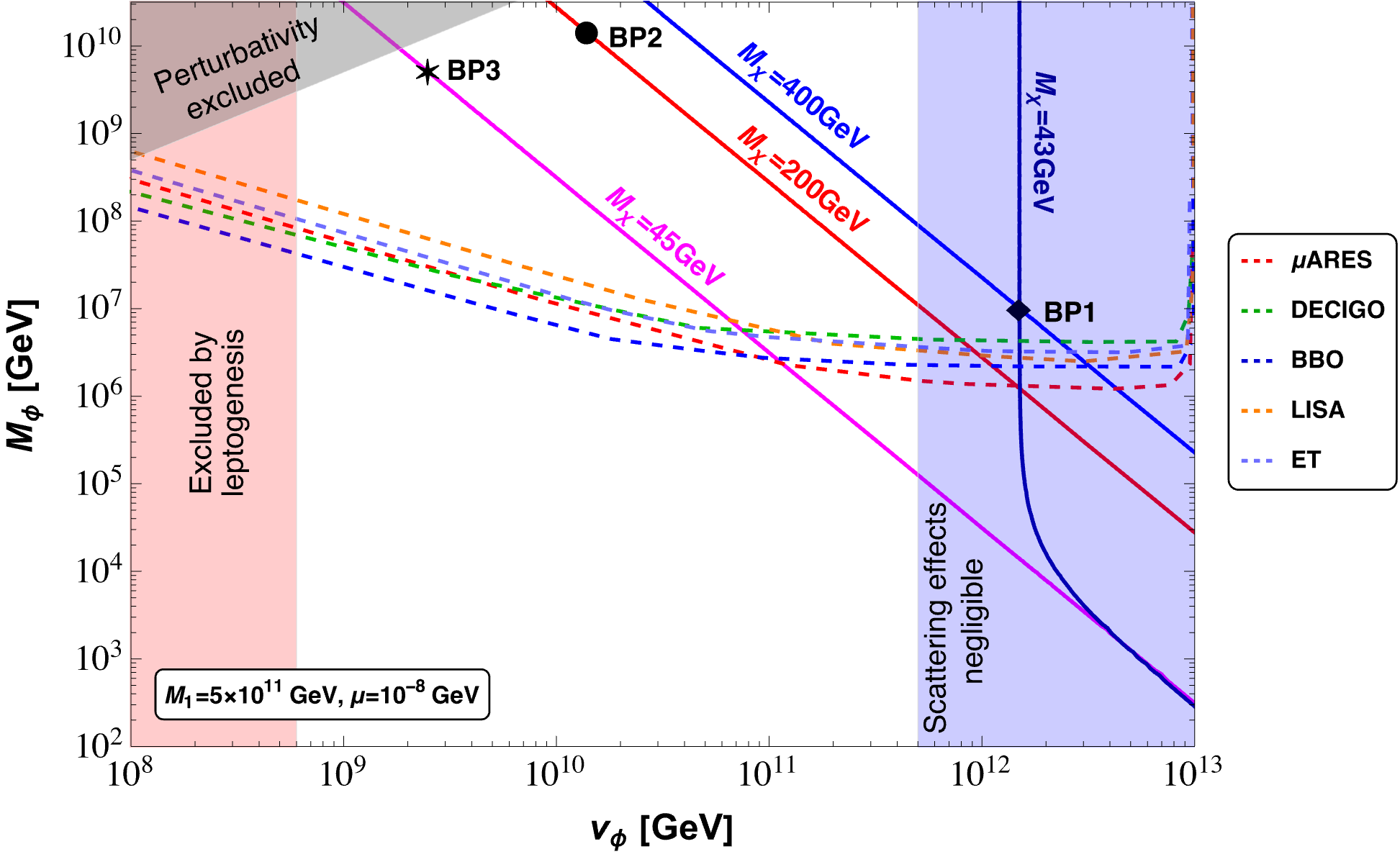}
\caption{Parameter space of the model to simultaneously explain neutrino masses, baryon asymmetry and dark matter abundance, for $M_1 = 5 \times 10^{11}$ GeV and $\mu = 10^{-8}$. The solid lines represent $m_\chi$ contours reproducing the correct relic abundance, whereas, the dashed lines signify the sensitivity curves of various gravitational wave detection experiments (current and future). In the blue region, $N \phi$ scatterings can be neglected while the gray and red shaded regions are excluded by perturbativity and leptogenesis respectively.}
\label{fig:Summary}
\end{figure}

In Fig.~\ref{fig:Summary}, we show the viable parameter space of our model which can be probed at current experiments. We show the sensitivity curves of various experiments in the $v_\phi - M_\phi$ plane, while $\mu$ is fixed at $10^{-8}$ GeV. The grey shaded region in the top right corner of the plot is excluded from the perturbativity of the scalar potential. 
The blue shaded region on the right corresponds to the high $v_\phi$ regime where the $N \phi$ scattering effects are negligible and canonical type-I leptogenesis is recovered, whereas, the red shaded region on the left side represents the region where one cannot produce a lepton asymmetry that can match observed baryon asymmetry of the universe. Note that these regions correspond to our choice of $M_1  = 5 \times 10^{11}$ GeV. For higher (small) values of $M_1$, we expect the regions to shift towards right (left), since for a fixed value of $v_\phi$, $\gamma_S$ increased with $M_1$, and scatterings will lead to strong washout (see Eq.~\eqref{eq:reactapprox}).

We also show the contours of DM mass that satisfy the observed DM relic abundance in the same place as well denote where our benchmark points lie in the $v_\phi - M_\phi$ plane. It can be seen that two contours corresponding to distinct $m_\chi$ pass through {\bf BP1}, due to production via two different channels $\phi \rightarrow \chi\chi$ and $H\rightarrow\chi\chi$ discussed above. Finally we also calculate the signal-to-noise ratio~\cite{Maggiore:1999vm,Allen:1997ad}
\begin{equation}
    \varrho=\left[n_{\mathrm{det}} t_{\mathrm{obs}} \int_{f_{\min }}^{f_{\max }} d f\left(\frac{\Omega_{\text {signal }}(f)}{\Omega_{\text {noise }}(f)}\right)^2\right]^{1 / 2} \; ,
    \label{eq:SNR}
\end{equation}
for different GW detectors like ET~\cite{Punturo:2010zz}, LISA~\cite{LISA:2017pwj}, DECIGO~\cite{Kawamura:2020pcg}, $\mu$Ares~\cite{Sesana:2019vho}, SKA~\cite{Janssen:2014dka} and THEIA~\cite{Garcia-Bellido:2021zgu}, following which we plot their individual sensitivity curves with $\rm{SNR}=10$ in Fig.~\ref{fig:Summary}. We see that the model has a viable parameter space that lies well within the current and future sensitivity curves of GW detection experiments.

\section{Summary and Conclusions}\label{sec:summary}

We explore a model where we connect the seesaw-I model to DM production 
and leptogenesis. The main ingredient was to think of a 
scalar mediator ($\Phi$) to append in the RHN-Higgs Yukawa 
term suppressed by a NP scale, as well as in the Yukawa term with 
a fermion DM, via imposing 
a $\mathcal{Z}_4$ symmetry. The scalar acquires a vev and breaks $\zf$ to a remnant $\mathcal{Z}_2$, which keeps the DM stable. 
The vev in turn generates the mass term for DM, as well as the 
required RHN-SM Yukawa term responsible for neutrino mass 
generation and leptogenesis. The choice of $\zf$ was minimal to 
keep all the possibilities on board, however, one can choose $\mathcal{Z}_6$ or $\mathcal{Z}_8$ as well. 

The phenomenology of the model mainly depends on the choice of $\zf$ breaking scale, $v_\phi$. This in turn is guided by whether the DM
is in the thermal bath or not. Following the absence of DM signal in current experiments, we choose the DM to be feebly coupled, which necessitates the $v_\phi$ to be large. While this ensures a natural 
agreement to neutrino mass generation via type-I seesaw and a larger 
parameter space in agreement with oscillation data, it also 
allows plenty of possibilities of DM production from the thermal bath
including that from the $\Phi$ decay, allowing it to saturate observe 
density.

More interestingly the connection with the scalar ($\Phi$) 
not only ensures a decay-driven leptogenesis, but additional scattering contribution 
to generate the CP asymmetry, with both the processes playing an important role in determining the final lepton asymmetry. We derive the CP asymmetry generated by such graphs including the loop graphs and derive the Boltzmann Equations including such terms. Part of this analysis serves to cater all such possibilities where scattering-driven contribution occurs. In our case, we see that these additional contributions effect in washout of the lepton asymmetry, and thus modify the usual contribution of vanilla leptogenesis, subject to the $\zf$ breaking scale.  

Spontaneous $\zf$ breaking can lead to domain wall problems, and that can be tackled via introducing soft explicit $\zf$ breaking terms which lifts the degenerate vacuum. This may cause the domain walls to collapse and annihilate in a radiation-dominated era resulting in gravitational wave generation, which we show to be within the reach of various future experimental sensitivities. Again, this pertains to 
the $\zf$ breaking scale, thus correlating all the possibilities together to be probed via gravitational wave signal detection.

\acknowledgments

We would like to thank P.K. Dhar for the useful discussions in the early stages of this
project. DV is supported by the “Generalitat Valenciana” through the GenT Excellence Program (CIDEGENT/2020/020). NM thanks the Council of Scientific \& Industrial Research (CSIR), Govt. of India for the junior research fellowship. RR acknowledges financial support from the STFC Consolidated Grant ST/T000775/1. We also want to thank Qaisar Shafi, Xin Wang, and Rinku Maji for the various fruitful discussion related to the GW from DWs.

\bibliographystyle{JHEP}
\bibliography{refs}

\end{document}